\documentclass[twocolumn,prd,nofootinbib]{revtex4-1}
\usepackage{graphicx}
\usepackage{amsmath}
\usepackage{hyperref}

\newcommand{\be}{\begin{equation}}
\newcommand{\ee}{\end{equation}}
\newcommand{\beq}{\begin{eqnarray}}
\newcommand{\eeq}{\end{eqnarray}}

\newcommand{\scale}{\left(\frac{a_F}{a}\right)}

\def\nus{\mathrel{{\nu_s}}}

\def\nur{\mathrel{{\nu_r}}}
\def\nua{\mathrel{{\nu_\alpha}}}

\def\ms{\mathrel{m_4}}
\def\mx{\mathrel{m_5}}

\def \lta {\mathrel{\vcenter{\hbox{$<$}\nointerlineskip\hbox{$\sim$}}}}
\def \gta {\mathrel{\vcenter{\hbox{$>$}\nointerlineskip\hbox{$\sim$}}}}

\def\t13{\mathrel{{\theta_{13}}}}
\def\y12{\mathrel{{\tan^2 \theta_{12}}}}
\def\c2{\mathrel{{\chi^2 }}}

\def\neff{{N_{\rm eff}}}
\def\neffc{{N^{\rm CMB}_{\rm eff}}}
\def\neffb{{N^{\rm BBN}_{\rm eff}}}
\def\dneff{\Delta N_{\rm eff}}

\def\yp{\mathrel{{Y_p}}}

\newcommand{\hide}[1]{{}}
\newcommand{\n}{neutrino}
\newcommand{\ns}{neutrinos}

\begin{document}
\title{Additional light sterile neutrinos and cosmology}
\author{Thomas D. Jacques$^{1,2}$}
\author{Lawrence M. Krauss$^{1,2}$}
\author{Cecilia Lunardini$^{2}$}

\affiliation{$^1$School of Earth and Space Exploration, Arizona State University, Tempe, Arizona, 85287-1404, USA}

\affiliation{$^2$Physics Department, Arizona State University, Tempe, Arizona, 85287-1404, USA}


\date{\today}

\begin{abstract}

Tantalizing cosmological and terrestrial evidence suggests the number of light neutrinos may be greater than 3, motivating a careful reexamination of cosmological bounds on extra light species.  Big bang nucleosynthesis constrains the number of relativistic neutrino species present during nucleosynthesis, $\neffb$, while measurements of the cosmic microwave background (CMB) angular power spectrum constrain the effective energy density in relativistic neutrinos at the time of matter-radiation equality, $\neffc$. There are a number of scenarios where new sterile neutrino species may have different contributions to $\Delta \neffb$ and $\Delta \neffc$, for masses  that may be relevant to reconciling cosmological constraints with various terrestrial claims of neutrino oscillations. We consider a scenario with two  sterile neutrinos and explore whether partial thermalization of the sterile states can ease the tension between cosmological constraints on $\neffb$ and terrestrial data. We then investigate the effect of a nonzero neutrino mass on their contribution to the radiation abundance, finding reductions in $\Delta\neffc$ of more than 5\% for neutrinos with masses above 0.5 eV. While the effects we investigate here could play a role, we nevertheless find that two additional light sterile neutrinos species cannot fit all the data at the 95$\%$ confidence level.

\end{abstract}

\maketitle

\section{Introduction}

Neutrinos are unique among the known elementary particles in that their properties have often been first, and in general more, constrained by astrophysical and cosmological limits than by direct laboratory measurements.  Already in the 1970s 
cosmological probes gave, first, a constraint on the \n\ mass based on estimates of the density of nonrelativistic matter in the Universe \cite{Cowsik:1972gh}, and then a constraint on the number of light neutrino species based on estimates of primordial helium production during big bang nucleosynthesis (BBN) \cite{Steigman:1977kc,1990ApJ...358...47K,k&k}.  
Today, both of these probes have reached impressive sensitivity, and have started  yielding some tantalizing suggestions of the possibility that extra neutrinos may be present in nature.

The radiation abundance in neutrinos and beyond-standard-model relativistic species is usually expressed as the effective number of relativistic species,
\begin{equation}
\neff = \frac{\rho_{\rm rel} - \rho_\gamma}{\rho_\nu^{\rm th}},
\end{equation}
where $\rho_{\nu}^{\rm th}=(7\pi^2/120)(4/11)^{4/3}T_\gamma^4$ is the energy density of one standard-model massless neutrino with a thermal distribution,  $\rho_\gamma $ is the energy density of photons,  and $\rho_{\rm rel}$ is total  energy density in relativistic particles.
In the standard model, by the time of BBN  only the three known \n\ species contribute to $\rho_{\rm rel}$, resulting in $\neff = 3.046$  \cite{mangano}. This is slightly larger than three due to reheating via $e^+e^-$ annihilation.  

Extra radiation beyond the standard model  (the so-called ``dark" radiation), would cause an excess (which we label $\Delta \neff $) above the standard model value of $\neff$. Although adding an extra light fermion could contribute $\Delta\neff =1$,  most generally $\neff$ is noninteger and varies with time, and depends on the physics at play. 
Specifically, lepton asymmetries \cite{Krauss:2010xg,Hannestad:2012ky,Mirizzi:2012we}, particle decay \cite{Ma:1999im,PalomaresRuiz:2005vf,Ichikawa:2007jv}, partial thermalization of new fermions \cite{Cirelli:2004cz,Melchiorri:2008gq,Boyanovsky:2006it}, the effect of a new MeV-scale particle on the active neutrino temperature  \cite{Ho:2012br,Boehm:2012gr}, nonthermal production of dark matter \cite{Hooper:2011aj}, and heavy sterile neutrinos can all lead to contributions to $\neff$ that are not integer and/or change with time. Therefore we can hope that probing $\Delta \neff$ precisely at different epochs -- namely, during BBN and at the formation of the CMB -- could discriminate between different models. 
%

Recent  measurements of $\neff$ have hinted at a value of $\neff > 3.046$ ($\Delta \neff > 0$). 
Constraints on $\neff$ can be derived from measurements of the primordial $^4$He mass fraction, $\yp \equiv \frac{4 n_{He4}}{n_n + n_p}$, at BBN, $T \sim 0.2$ MeV.
Izotov and Thuan \cite{izotov} find $Y_p = 0.2565 \pm 0.0010$(stat.)$\pm  0.0050$(syst.), giving  $\neffb = 3.68^{+0.80}_{-0.70}$ or $ \neffb =3.80^{+0.80}_{-0.70}$, each at 2$\sigma$, depending on the choice of the neutron lifetime, and assuming no lepton asymmetry. These are both more than $1\sigma$ from the standard model value. Other recent estimates of $Y_p$ \cite{skillman,Aver:2011bw} and various analyses of $\neff$ at BBN, e.g.~\cite{Mangano:2011ar,Nollett:2011aa,Hamann:2011ge},  give for the most part central values more than 1$\sigma$ above 3.

CMB measurements constrain the neutrino energy density in two ways.  First, a measurement of the damping tail of the angular power spectrum on small scales (large $l$) is a probe of the energy density in light neutrinos which can free stream during structure formation. Next, measurements of the angular power spectrum at larger scales near the Doppler peak can be used to constrain the redshift of matter-radiation equality.
With independent measurements of the total matter abundance, this can also constrain the radiation abundance at the time of matter-radiation equality.
Planck reports a value of $\neffc = 3.30 \pm 0.27$, consistent with the standard model at the $1\sigma$ level \cite{Ade:2013lta}. The South Pole Telescope suggests  a somewhat high value, $\neffc = 3.71 \pm 0.35$ \cite{Hou:2012xq}. WMAP 9 also reports a value around 2$\sigma$ higher than the standard model value, $\neffc = 3.84 \pm 0.4$ \cite{Hinshaw:2012fq}. In contrast with this, the Atacama Cosmology Telescope (ACT) finds a significantly smaller value, $\neffc = 2.78 \pm 0.55$  when using CMB data alone, although this value shifts to $\neffc = 3.52 \pm 0.39$ when baryon acoustic oscillation and Hubble parameter measurements are included \cite{Sievers:2013wk}.

Interestingly, bounds from terrestrial searches for new physics on the masses and couplings of new particles  invariably result in constraints on their contribution to the cosmological radiation, providing indirect constraints on $\dneff$.
Of particular interest are the recent hints of a fourth, sterile \n\ species from  reactor \n\ experiments \cite{Mueller:2011nm,Mention:2011rk},  calibration data from gallium-based solar \n\ detectors \cite{Giunti:2010zu,Kaether:2010ag,Abdurashitov:2009tn}, and the Short Baseline (SBL) \n\ beam  experiments LSND \cite{Aguilar:2001ty} and MiniBooNE \cite{AguilarArevalo:2008rc,AguilarArevalo:2009xn,AguilarArevalo:2010wv,AguilarArevalo:2012va} which search for $\bar \nu_\mu \rightarrow \bar \nu_e$ and $\nu_\mu \rightarrow \nu_e$ oscillations.  All these generally support the existence of at least one sterile \n\  with mass $\sim$ 0.1 -- 1  eV. 
This \n\ would be populated in the early Universe via an interplay of oscillations and scattering, thus increasing $\neff$. 

Although the possibility of extra radiation due to sterile \ns\ seems to be substantiated at the general level, detailed analyses of the data reveal tensions between data sets and leave open the question of what scenario is most favored overall.  
MiniBooNE observes a difference between the muon neutrino and antineutrino disappearance rates, hinting at \emph{CP} violating effects. The latest measurements by MiniBooNE show less tension between their neutrino/antineutrino results,  although the $3+2$ scenario still provides a better fit to the data  \cite{AguilarArevalo:2012va}. The simplest explanation for this is the existence of two sterile neutrinos families, although other data does not easily accommodate that possibility  and the improvement in the global fit to the data may simply be due to the additional free parameters in a $3+2$ model \cite{Maltoni:2007zf,Kopp:2011qd,Giunti:2011gz,Abazajian:2012ys,Conrad:2012qt}.  Fits of a $3+2$ model to cosmological data \cite{Joudaki:2012uk} and combined fits of SBL and cosmological data \cite{Archidiacono:2012ri} have found further tension when the sterile neutrinos are fully thermalized, with the level of tension depending on exactly which data sets are considered.

Whilst the cosmological data appear to rule out $\neff =5$, multiple sterile neutrinos can still be accommodated if one or more of them are not fully thermalized at the time of BBN. The degree of thermalization depends on the sterile neutrino mass and mixing parameters, as constrained by SBL data. 
Neutrino density evolution and partial thermalization in a $3+2$ scheme has been studied by Melchiorri \emph{et al.} \cite{Melchiorri:2008gq}, finding tension both between the various terrestrial data sets, and between terrestrial and cosmological data. 
Since then, there have been substantial improvements in cosmological measurements and experimental results.

Considering that there is evidence from multiple sources that perhaps additional light neutrinos exist, and scenarios with two sterile neutrinos have been proposed as a way to explain the MiniBooNE results, it is important to reexamine cosmological constraints with a more careful eye.  With this goal in mind we have explored both partial thermalization of the sterile species at BBN and the effects that small neutrino masses will have on the relativistic neutrino fraction at the time of matter-radiation equality.  As we will show, $\neff$ at BBN and CMB can be quite different for light neutrinos, but even incorporating this fact, and pushing all constraints to their $2\sigma$ level, cosmology can still not accommodate models with two such neutrinos. 

\section{Partial Thermalization in a $3+2$ Scenario and BBN Constraints\label{partialtherm}}

We consider a scenario with two sterile \ns, and study the sensitivity of their thermalization efficiency to their masses and mixings.  Specifically, we denote the two sterile \n\ flavors as $\nu_s$ and $\nu_r$, and the corresponding mass eigenstates as $\ms,\,\mx$, such that the mixing matrix $U$ has entries  $U_{s 4} \simeq U_{r 5} \simeq 1$, and the hierarchy $m_5 > m_4 \gg m_j$ ($j=1,2,3$) holds.  For simplicity, we also assume $U_{\tau 4} = U_{\tau 5} = 0$, so that our results for $\neffb$ only depend on  
$U_{e4},\, U_{\mu 4}, \, U_{e5}, \, U_{\mu 5}$.  We have verified that the complex phase $\eta = {\rm Arg}(U^*_{e4} \, U_{\mu 4} \, U_{e 5} \, U^*_{\mu 5})$ does not affect the degree of thermalization, therefore we simplify the notation by considering $U$ to be real. 
%

In the density matrix formalism the differential equations governing evolution of the neutrino density are

\begin{equation}
\dot{\rho} = \mathcal{H} \rho - \rho \mathcal{H}^\dagger = i [H_m + V_{\rm eff} \, , \, \rho ] - \{ \frac{\Gamma}{2} , (\rho - \rho_{eq})\},
\label{ev_eqn1}
\end{equation}
where $\rho$ is the $5\times 5$ neutrino density matrix in the flavor basis with diagonal entries corresponding to physical densities, $\mathcal{H}$ is the full Hamiltonian, $H_m = U \, H_0 \, U^\dagger$ is a rotation of the  free neutrino Hamiltonian in the mass basis $H_0 = {\rm diag}(E_1,E_2,E_3,E_4,E_5)$, and $\rho_{eq}$ is the density matrix at thermal equilibrium, $\rho_{eq} = I \left(1/\left(1+e^{E/T}\right)\right)$.
Equation~(\ref{ev_eqn1})  can be expressed as
\begin{equation}
\left(\frac{\partial\rho}{\partial T}\right)_\frac{E}{T} = -\frac{1}{H T}\left(i [H_m + V_{\rm eff} \, , \, \rho ] - \{ \frac{\Gamma}{2} , (\rho - \rho_{eq})\}\right),
\label{ev_eqn}
\end{equation}
using the approximation $\dot{T} \simeq H T$, where $H = \sqrt{\frac{4 \pi^3 g^\ast}{45}} \frac{T^2}{M_{\rm pl}}$ is the Hubble parameter, $M_{\rm pl}$ is the Planck mass and $g^\ast$ is the effective number of relativistic degrees of freedom.
Since the full Hamiltonian is complex, the equation decomposes into a coherent commutator, and an anticommutator which describes loss of coherence. 
$V_{\rm eff} = I (V_e, V_\mu, V_\tau, 0, 0)$ describes the effects of matter on the coherent part of the neutrino evolution. 
For zero lepton asymmetry,
\begin{equation}
V_{\alpha} = - A_\alpha \frac{2 \sqrt{2} \zeta (3)}{\pi^2} \frac{G_F T^4 p}{m_W^2},
\label{potential}
\end{equation}
 where $A_e = 17$ and $A_{\mu,\tau} = 4.9$.   Here the negligible contribution of the baryon asymmetry is omitted. For simplicity, we do not consider the richer phenomenology that arises if a large lepton asymmetry exists.
The vector $\Gamma = I (\Gamma_e, \Gamma_\mu, \Gamma_\tau, 0, 0)$ encodes decoherence and damping due to  collisions with the background medium, 
\begin{equation}
\Gamma_\alpha \simeq y_\alpha \frac{180 \zeta (3) }{7 \pi ^4} G_F^2 T^4 p,
\label{collisionrate}
\end{equation}
with $y_e = 3.6, \, y_{\mu,\tau} = 2.5$.

Before discussing the full numerical solution of Eq.~(\ref{ev_eqn}), we start with an approximate analytical solution for guidance in understanding the physics. Briefly (see the Appendix for more details), the problem can be approximately reduced to two independent equations, each describing the population of one of the sterile species. For each sterile neutrino, (we use $\nu_s$ as an example in the following expressions), one can approximately use two independent oscillation channels, $\nu_e \rightarrow \nu_s$ and $\nu_\mu \rightarrow \nu_s$. For each  channel, the effective mixing angle in vacuum is given by
\begin{equation}
\sin^2 2\theta_{\alpha s}\simeq   4  U^2_{\alpha 4} U^2_{s 4} \simeq 4  U^2_{\alpha 4}  ~,\hskip 0.5truecm  \alpha=e,\mu~,
\end{equation}
while in-medium the mixing is suppressed according to the expression
\begin{eqnarray}
\sin^2 2\theta_m &\simeq& \frac{\sin^2 2\theta_{\alpha s}}{(1- b_\alpha(p,T))^2},\\
b_\alpha(p,T) &=& \frac{2 E \,V_\alpha}{\Delta m^2}~.
\end{eqnarray}
One can then solve the evolution equation for $f_s$, the phase space distribution of $\nu_s$, in terms of the interplay of oscillations and collisions. If $f_s/f_\alpha$ is energy independent (i.e., a constant), and the mixing $\theta_m$ is well in the vacuum limit at the freeze-out epoch, we find the contribution of $\nu_s$ to $\neffb$ to be
\begin{align}
&\Delta N^{\rm BBN}_{\rm eff, s} = \frac{f_s}{f_\alpha} \simeq  \nonumber\\
&1 - \exp \left[ \frac{-  2.06\times 10^3 }{\sqrt{g^\ast}}   \left(\frac{m_4 }{eV} \right)    \left( U^2_{e4} +  1.29 U^2_{\mu 4}  \right) \right] .
\label{analytic1}
\end{align}
A similar expression holds for $\nu_r$, with the substitutions $U^2_{\alpha 4} \rightarrow U^2_{\alpha 5}$ and $\ms \rightarrow \mx$.  Ultimately, the total contribution of the two sterile states to $\neff$ is given by
\begin{align}
\Delta N^{\rm BBN}_{\rm eff} \simeq \Delta N^{\rm BBN}_{\rm eff, s} + \Delta N^{\rm BBN}_{\rm eff, r}~. 
\label{analytic2}
\end{align}
As expected, a sterile species is more populated at the time of BBN if oscillations are more efficient, i.e. for larger mixing (larger oscillation amplitude) and larger mass squared splitting relative to the active species, which means smaller oscillation length and therefore a higher probability of flavor conversion between two successive collisions. 
We also stress (see the Appendix) that $\Delta N^{\rm BBN}_{\rm eff, s}$ is larger for a larger collision rate; indeed, collisions favor the growth of the sterile population toward equilibrium \cite{Enqvist:1991qj}  and in the limit of no collisions (oscillations only), we would have $\frac{f_s}{f_\alpha}  \leq (\sin^2 2\theta_{e s} + \sin^2 2\theta_{\mu s})/2$, where the right-hand side is the sum of the average vacuum oscillation probabilities in the two channels. Note that the production of $\nu_s,\nu_r$ via oscillation from $\nu_\mu$ is more efficient, because for the $\nu_\mu-\nu_s$ system the mixing angle is less suppressed by the thermal potential ($|V_\mu| <| V_e|$).  

Let us now discuss the numerical solution.  We follow the technique from Melchiorri \emph{et al.} \cite{Melchiorri:2008gq}, numerically evolving the neutrino densities from temperatures of 100 MeV down to 1 MeV. 
To simplify the resultant set of differential equations, we assume a monochromatic neutrino energy distribution, with $E_\nu \simeq 3.15 T$, rather than use the full spectrum of the \ns. This simplification has little effect on the density evolution \cite{Melchiorri:2008gq}. 

\begin{figure}[htbp]
  \centering \includegraphics[width=0.45\textwidth]{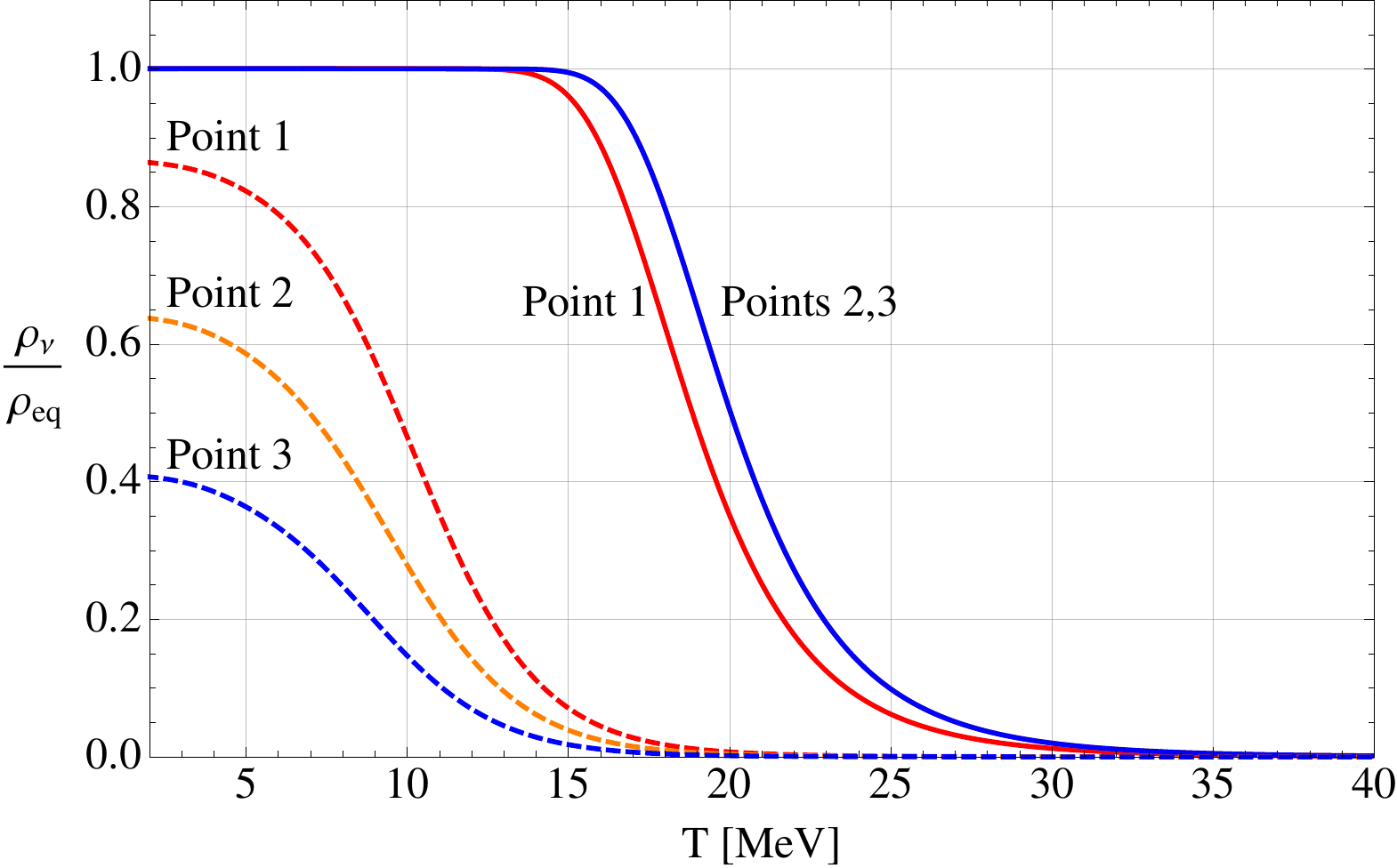}
 \caption{Sterile neutrino density evolution as a fraction of the thermal density $\rho_{\rm eq}$, for the masses and mixing angles listed in  Table~\ref{parameters}. Dashed lines are for $\nu_{s}$, solid lines are for $\nu_{r}$. }
\label{results1}
\end{figure}

%
We begin by running a loose scan across the allowed parameter space with the goal of finding reference points that minimize $\neff$ whilst also keeping $\ms$ and $\mx$ as low as possible, due to strong cosmological constraints on the sum of the neutrino masses.
Constraints on the masses and mixing parameters are from SBL data \cite{Giunti:2011gz}.
As shown in Eq.~(\ref{analytic1}), the contribution of each sterile neutrino to $\neff$ is smallest when the two mixing matrix elements for that neutrino are minimized.  However, SBL constraints on the product of the four mixing matrix elements prevent all four elements from being small.
By definition, $\mx > \ms$, and so it can be seen from Eq.~(\ref{analytic1}) that $\nu_r$ will have a larger contribution to $\neffb$ than $\nu_s$ for comparable mixing angles. For this reason, we focus on minimizing the thermalization of $\nu_s$.
Point 1 in Table~\ref{parameters} is chosen to correspond to the minimum values of  $U_{e4}, \, U_{\mu 4}$ still allowed within 2$\sigma$. 
$U_{e5}, \, U_{\mu 5}$ are chosen to be as small as possible while satisfying constraints on  $4 | U_{e4} \, U_{\mu 4} \, U_{e5} \, U_{\mu 5}|$. 
$\ms, \, \mx$ are chosen to be as low as possible while still allowed by our choices of $U_{ij}$. 
This point does indeed lead to incomplete thermalization of $\nu_s$, with $\Delta\neffb = 1.86$, as shown in Fig.~\ref{results1}, although this is still outside the 2$\sigma$ allowed range from Izotov and Thuan \cite{izotov}, $\neffb = 3.80^{+0.8}_{-0.7}$.
In this region, the degree of thermalization is quite sensitive to $U_{e4}, \, U_{\mu 4}$. As an illustration, Fig.~\ref{results1} also shows the density evolution for two additional points in parameter space, labeled points 2 and 3 in Table~\ref{parameters}, where $U_{e4}, \, U_{\mu 4}$ are pushed to even lower values, outside of the 2$\sigma$ allowed region, but still within the 99 \% C.L. allowed region. $\ms, \, \mx$ are kept fixed, and $U_{e5}, \, U_{\mu 5}$ are chosen to keep $4 | U_{e4} \, U_{\mu 4} \, U_{e5} \, U_{\mu 5}|$ as close to the $2\sigma$ allowed region as possible while still remaining within the $2\sigma$ allowed region themselves. At points 2 and 3, $\Delta\neffb$ is safely within the Izotov and Thuan 2$\sigma$ allowed range. 
%

\begin{table}
  \centering 
  \begin{tabular}{c|c|c|c|c|c|c|c|c|c}
\hline 
 		 & $U_{e4}$	& $ U_{\mu 4}$	& $U_{e5}$	& $U_{\mu 5}$	&$\ms$	& $\mx$	& $\dneff$& $\dneff$	 & $\sum m_{\nu_s}^{\rm eff}$\\ 
		 &			&			&			&			& (eV)	& (eV)	& (BBN)	&  ($z_{eq}$) 	& (eV)	\\ \hline \hline
   Pt. 1   & 0.055		& 0.034		& 0.13		& 0.13		& 0.6		& 0.9		& $1.86$	& 1.68 & 1.31		\\ \hline
   Pt. 2   & 0.040		& 0.025		& 0.17		& 0.17		& 0.6		& 0.9		& $1.63$	& 1.47 & 1.18		\\ \hline
   Pt. 3   & 0.030		& 0.016		& 0.17		& 0.17		& 0.6		& 0.9		& $1.40$ &1.25	 & 1.05	\\ \hline
\end{tabular}
  \caption{ Mass, mixing parameters, results for $\dneff$ at BBN and $z_{\rm eq}$, and effective mass sum for the three sample points discussed in the text. The derivation of the final three columns is discussed in Secs.~\ref{partialtherm} and  \ref{partialCMB}.\label{parameters} }
\end{table} 

\section{Partial Thermalization and Partially Relativistic Neutrinos in a $3+2$ Scenario: CMB Constraints\label{partialCMB}}

Both the position of the lowest peaks in CMB angular power spectrum and the damping tail at high multipole moments, $l$, are sensitive to the redshift at matter-radiation equality, and hence to the energy density of relativistic neutrinos at that time.     
The latter is usually expressed as a constraint on the sum of the neutrino masses,
\begin{equation}
\label{Omega2MassSum}
\sum m_\nu = 94 {\rm eV} (\Omega_{\nu,m} h^2),
\end{equation}
where $\Omega$ is the density as a fraction of the critical density of the Universe, $\Omega = \rho/\rho_c$, and $\Omega_{\nu,m}$ is the neutrino contribution to the matter abundance $\Omega_m$.  It is important to note that in Eq.~(\ref{Omega2MassSum}) it is assumed that each species is fully thermalized, by assuming that for each species, $\rho_\nu^{\rm non-rel} = m_\nu n_\nu^{\rm th}$, with $n_\nu^{\rm th}$ from Eq.~(\ref{number-density}). Constraints on the sterile neutrino mass are really constraints on the product $m_\nu n_\nu$, and if a sterile neutrino does not undergo full thermalization, then it contributes $m_{\rm eff} = m_\nu \frac{n_\nu}{n_\nu^{\rm th}}$ to constraints on $\sum m_\nu$. For the partially thermalized $m_\nu = 0.6$ eV neutrino we consider in Table~\ref{parameters}, the phase space distribution is approximately a scaled Fermi-Dirac distribution as shown in the Appendix, and $\frac{n_\nu}{n_\nu^{\rm th}} = \Delta N_{\rm eff}^{\rm CMB}$. In Table~\ref{parameters} we show the effective $\sum m_\nu$ for the three points we consider.

In addition, many upper limits on the sum of the neutrino masses assume the standard model value of $\neff$, and so do not directly apply to sterile neutrinos; however, a number of groups have constrained the $\neff - \sum m_\nu$ plane using various combinations of measurements of the CMB, Hubble constant, baryon acoustic oscillations, and galaxy clusters  \cite{RiemerSorensen:2012ve,Giusarma:2012ph,Wang:2012vh,Benson:2011ut}.   Other analyses perform global fits to the cosmological data, including the possibility of one or two massive and fully thermalized neutrinos which contribute a full $\dneff =1$ each, and additional massless species with noninteger contributions to $\dneff$ \cite{Hamann:2011ge,Joudaki:2012uk}.  For the value of $\neff$ we are interested in, detailed below,  $\sum m_\nu \gtrsim 0.7$ eV is excluded at the 95\% confidence level, which is inconsistent with the values shown in Table~\ref{parameters} for all three points.  (Tension between SBL and cosmological data is also discussed in Ref.~\cite{Joudaki:2012uk} in the context of two fully thermalized neutrinos.)

There is another, equally important factor that can affect the value of $\neff$ that should be utilized when applying cosmological constraints: the fact that neutrinos may not be fully relativistic at the time of matter-radiation equality. 

The standard model neutrino temperature at matter-radiation equality is  $T_\nu = 0.55$ eV \cite{Hinshaw:2012fq}.\footnote{
For massive neutrinos that are not in a thermal distribution, the neutrino temperature $T_\nu$ is not a meaningful physical quantity. However, the equivalent temperature of a massless neutrino is used throughout this work, as it is still valid as a convenient measure of time. $T_\nu = \left( \frac{4}{11}\right)^{\frac{1}{3}} \, T_\gamma$,  $T_\gamma = T_0 (1+z)$, and the scale factor $a = \frac{1}{1+z}$.
}
$\neff$ as derived from  the CMB measures the relativistic energy density at the time of matter-radiation equality. 
A neutrino with $m_\nu \sim \mathcal{O}(1\,{\rm eV})$ will not be entirely relativistic at this time, and so will contribute  $\Delta\neff \lesssim 1$. Constraints on $\neff$ are continually getting tighter, and so this can be an important effect for sterile neutrinos towards the top of the allowed mass range.
A similar effect was considered in Ref.~\cite{Dodelson:2005tp}, where the scale factor at matter-radiation equality was related to the mass and energy density of a sterile neutrino. 

The pressure density provides a convenient measure of how relativistic a particle is at any given temperature, with
 $P = \rho / 3$ for fully relativistic particles and $P=0$ for nonrelativisitc particles.  As the sterile neutrinos become less relativistic, their pressure drops below $\rho/3$, and the relativistic fraction of their energy density drops accordingly, 
\begin{equation}
\rho_\nu^{\rm rel} = \rho_\nu \left(\frac{P_\nu}{\rho_\nu} \middle/ \frac{1}{3} \right),
\end{equation}
using $P^{\rm th}_\nu/\rho^{\rm th}_\nu = 1/3$ when $m=0$. With this, the  effective number of relativistic degrees of freedom can  be expressed as 
\begin{equation}
\neff = \frac{ \rho_\nu^{\rm rel} }{ \rho_{\nu,m=0}^{\rm th}}=  \frac{P_\nu}{P_{\nu,m=0}^{\rm th}}.
\label{neff}
\end{equation}  

\begin{figure}[htbp]
  \centering \includegraphics[width=0.45\textwidth]{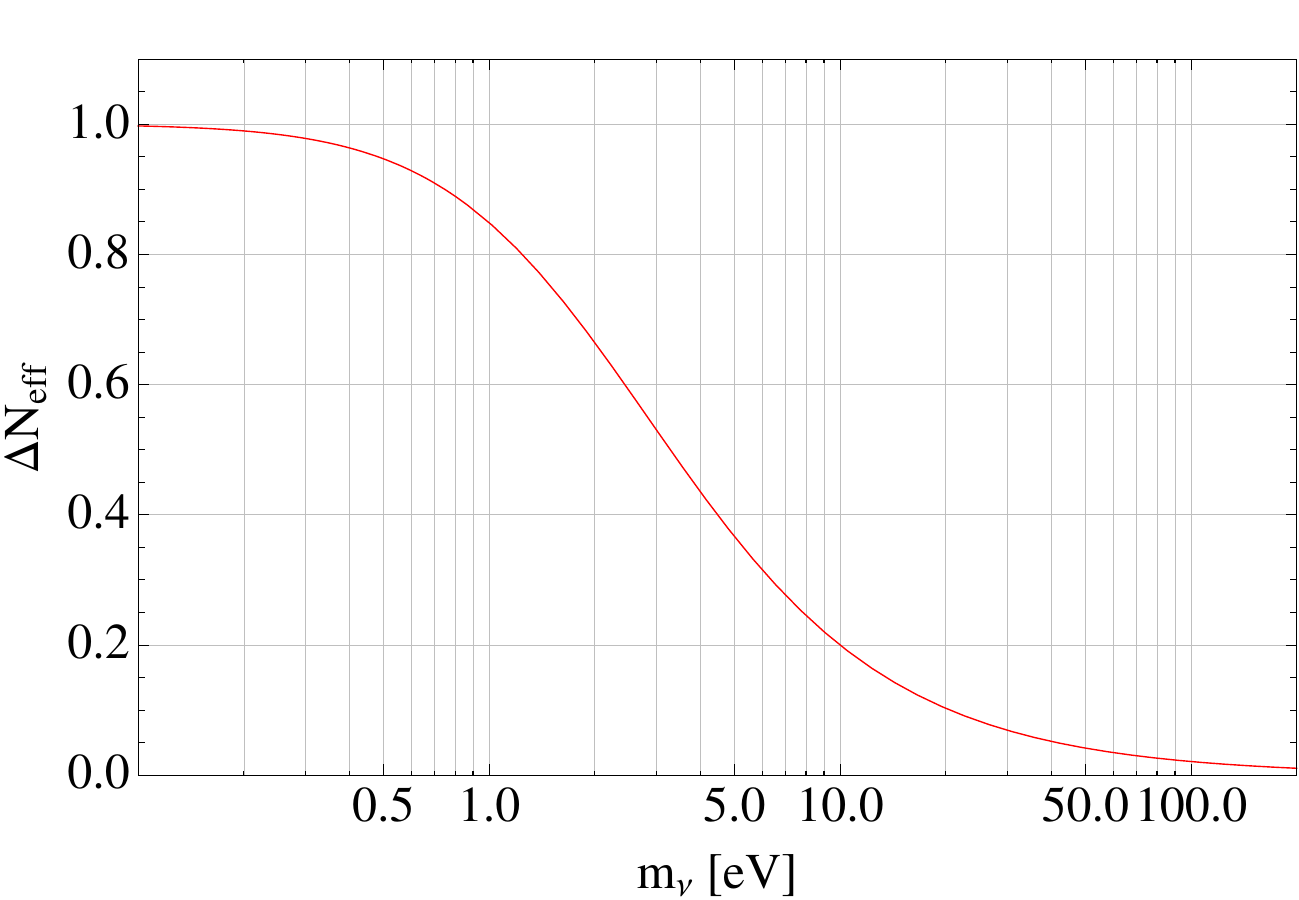}
 \caption{Contribution of one thermalized massive sterile neutrino to $\neff$ at the time of matter-radiation equality. If the sterile neutrino has an approximately Fermi-Dirac distribution, this is equivalent to $\Delta N_{\rm eff}^{z_{\rm eq}} / \Delta N_{\rm eff}^{\rm BBN}$, i.e. if the sterile neutrino is not fully thermalized and $\Delta N_{\rm eff}^{\rm BBN} < 1$, then $\Delta N_{\rm eff}^{z_{\rm eq}}$ will be reduced accordingly.}
\label{fig:Nvsm}
\end{figure}

\begin{figure}[htbp]
  \centering \includegraphics[width=0.45\textwidth]{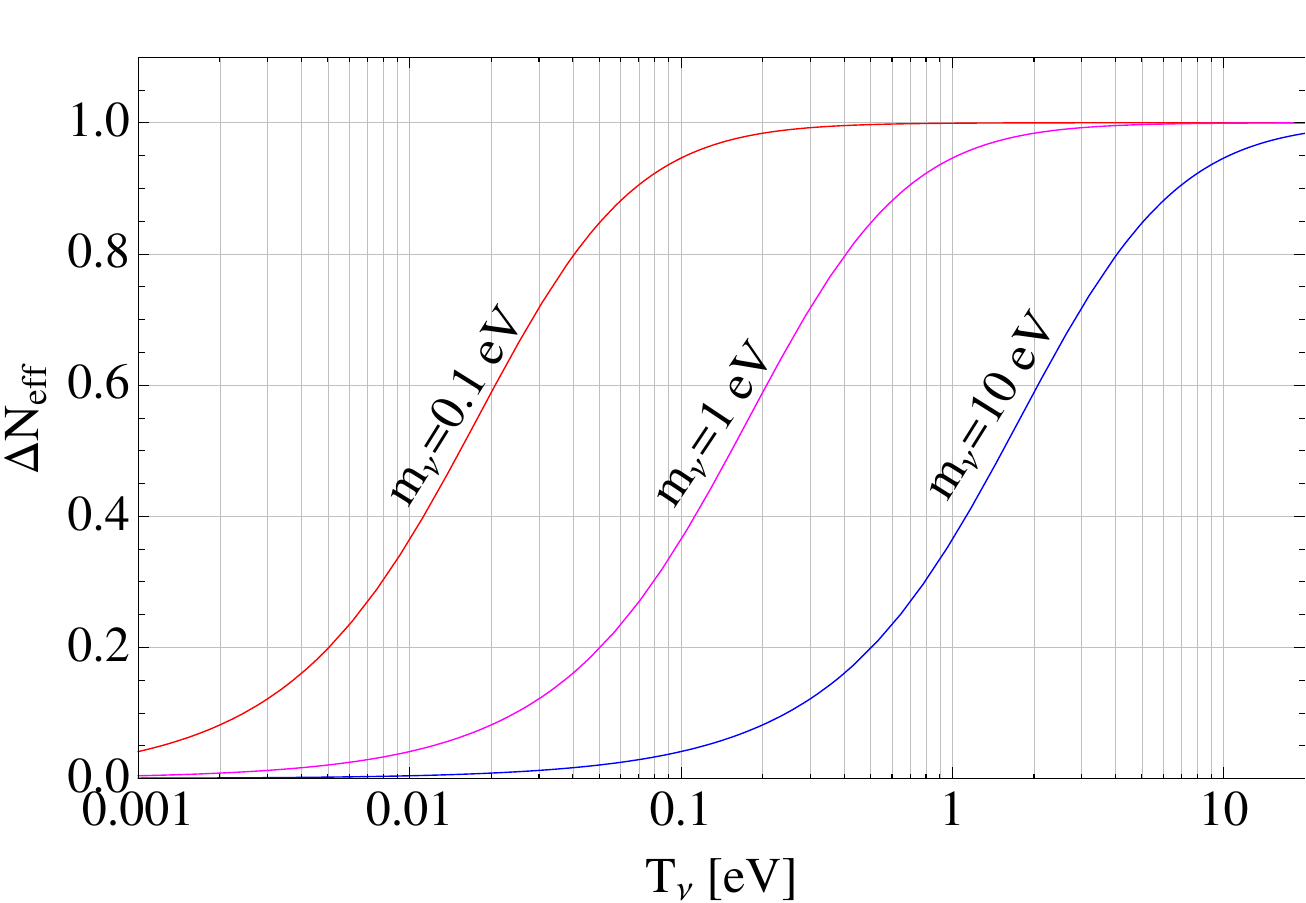}
 \caption{Contribution of one massive sterile neutrino to $\neff$ as a function of the equivalent temperature of a massless neutrino, $T_\nu$. At matter-radiation equality, $T_\nu = 0.55$ eV \cite{Hinshaw:2012fq}.}
\label{fig:NvsT}
\end{figure}

The number, pressure and energy densities are given by the standard formulas,
\begin{eqnarray}
n_\nu&=& \frac{g}{2 \pi^2} \int dp \, p^2 \, f_\nu(p),  \label{number-density}\\
P_\nu &=& \frac{g}{2 \pi^2} \int dp \frac{p^4}{3 E} f_\nu(p), \label{pressure-density} \\
\rho_\nu&=& \frac{g}{2 \pi^2} \int dp E \, p^2 f_\nu(p),  \label{energy-density}
\label{pressure}
\end{eqnarray}
where $g$ counts the number of helicity states, $f_\nu(p)$ is the neutrino phase-space distribution, and $p=|\vec{p}|$.
When the neutrinos are in thermal equilibrium, they follow a Fermi-Dirac distribution,
\begin{equation}
f_\nu(p) = \frac{1}{1+\exp[\frac{E}{T}]},
\end{equation}
when the chemical potential is zero.
After freeze-out at $T_F\sim 2$ MeV, the comoving number density must be conserved:
\begin{eqnarray}
n_\nu(p,T) & = & \scale^3 n_\nu(p_F, T_F)\nonumber\\
& = &\scale^3  \frac{g}{2 \pi^2} \int dp_F p_F^2 \frac{1}{1+\exp[\frac{E_F}{T_F}]},
\end{eqnarray}
where $a$ is the scale factor, and subscript $F$ denotes the value at freeze-out.
Neutrino momentum redshifts as $p = \scale \, p_F$, and so the neutrino number density after freeze-out is
\begin{widetext}
\begin{eqnarray}
n_\nu(p,T) &=& \scale^3  \frac{g}{2 \pi^2} \int \left( \frac{a}{a_F} d p\right) \, \left(\frac{a}{a_F} p\right)^2 \frac{1}{1+\exp\left[\frac{\sqrt{\left(\frac{a}{a_F} p\right)^2 + m^2}}{T_F}\right]}\nonumber\\
&=&\frac{g}{2 \pi^2} \int d p \, p^2 \frac{1}{1+\exp\left[\frac{\sqrt{\left(\frac{a}{a_F} p\right)^2 + m^2}}{T_F}\right]}.
\end{eqnarray}
\end{widetext}
Comparing this with Eq.~(\ref{number-density}) and using $T= T_0 / a $,  we have
\begin{equation}
f_\nu(p) = \left(1+\exp\left[ \frac{\sqrt{ \left(\frac{T_F\,p}{T}\right)^2+m^2} }{ T_{F}}\right] \right)^{-1}
\label{fnu}
\end{equation}
after freeze-out. This reduces to the standard expression for both relativistic and nonrelativistic particles. $\neff$ can then be found using Eq.~(\ref{neff}) with Eqs.~(\ref{pressure-density}) and (\ref{fnu}).
Since the comoving number density is conserved from the time when the neutrinos were entirely relativistic, it follows that the number density $dn/dp$ must be independent of mass. Thus, the total neutrino energy density $d \rho/dp = \sqrt{p^2+m^2} d n / dp$ will be larger for particles with larger mass. This is compensated for by the fact that the pressure is smaller for massive neutrinos, leading massive sterile neutrinos to have a smaller contribution to $\neff$ as shown in Figs.~\ref{fig:Nvsm} and \ref{fig:NvsT}. 

We use this relation to determine the relevant values of $\neff$  at matter-radiation equilibrium for the $3+2$ sterile neutrino models discussed earlier to explore whether they may be consistent with both SBL and cosmological bounds, and report the results in Table \ref{parameters}.\footnote{
Reference~\cite{Birrell:2012gg} performed a similar calculation of the effect of neutrino mass on $\neff$. Whilst our expression for the neutrino phase-space distribution, Eq.~(\ref{fnu}), agrees with their Eq.~(8), we reach a different conclusion regarding the effect on $\neff$, which measures the energy density in \emph{relativistic} neutrinos, rather than the total neutrino energy density.
}
 The use of Eq.~(\ref{neff})  to determine $\neff$ requires knowledge of the neutrino phase-space distribution at decoupling, and we demonstrate in the Appendix that our three points satisfy the conditions required for the phase-space distribution to be approximated by a Fermi-Dirac distribution scaled by a constant. With this approximation, and using Eq.~(\ref{neff}) with Eqs.~(\ref{pressure-density}) and (\ref{fnu}), the contributions to $\neff$ at $z_{\rm eq}$ are $\Delta\neff = (1.68, \, 1.47, \, 1.25)$ at points 1, 2, and 3 respectively.  This leads to some easing of the tension between SBL data and CMB constraints on $\neff$.

\section{Discussion and Conclusions\label{discussion}} 

We have discussed the cosmological generation and evolution of a population of two sterile \ns, with masses and mixings motivated by the recent SBL data. Specifically, we calculated the contribution of these extra \n\ species to $\neff$ at BBN and CMB epochs.  We focused on the region of the parameter space where the sterile \ns\ are produced with less than thermal abundance ($\dneff<2$), so that the tension with   BBN and CMB measurements is eased, compared with the case of two fully thermalized species.  We find points at the limit of the region of parameter space allowed by the SBL data where the heaviest sterile state is fully thermalized, while the second is produced with abundance as low as $\sim 40\%$ of the thermal abundance.

Whilst it is possible --- with the maximum suppression of $\neff$ due to partial thermalization ---  to find points in parameter space marginally compatible with BBN constraints ($\neffb \lta 4.6$ at 2$\sigma$),  the tension with BBN data overall remains. 

Interestingly, if  SBL-favored sterile \ns\ really are the origin of $\neff>3$, we expect their contribution to $\neff$ at $z_{\rm eq}$ --- relevant for CMB constraints --- to be lower than that at BBN epoch, due to their being only moderately (partially) relativistic at $z_{eq}$,  with a difference $\neffb - \neffc$ on the order of $10\%$ or less.  In principle, this feature would allow us to distinguish the sterile \n\ hypothesis from other possible origins of an excess of radiation. Future measurements of $\neffc$ could approach or reach this level of precision \cite{Joudaki:2011nw}.
We also note that while the mass-induced suppression works to ease the tension with the CMB data somewhat, it comes with a  price: the $\sim 1 $ eV masses of the sterile states would increase the sum of the \n\ masses to  $\sum m_\nu \gtrsim 1$ eV, which is disfavored by CMB bounds on this quantity. 
 
Summing up, we find that even with the suppression effects due to partial thermalization and partially relativistic masses, 
 two additional sterile neutrinos in a mass range that might explain SBL neutrino data appear to be inconsistent with cosmological bounds coming from BBN and CMB measurements.

It has recently been questioned whether the SBL data from MiniBooNE actually favor two sterile neutrinos  \cite{AguilarArevalo:2012va}. If this requirement is relaxed then the results we derive will be particularly relevant to constrain models with one extra neutrino.   Alternatively, new physics that might resolve these inconsistencies include altering microphysics or altering cosmology.  An example of the former includes introducing a lepton asymmetry in addition to the two sterile neutrinos \cite{FV,DiBari:1999ha,afp}, which can reduce the sterile neutrino abundance and distort the phase-space distribution \cite{Hannestad:2012ky,Mirizzi:2012we}.  As an example of the latter, Hamann \emph{et al.}~\cite{Hamann:2011ge} find that models with one fully thermalized eV-scale sterile neutrino and additional massless degrees of freedom can provide a better fit to a wide range of cosmological data than standard $\Lambda$CDM, if the dark energy equation of state parameter is free to be $w < -1$.  Nevertheless this requires a contribution to $\neff$ from massless sterile states of $\dneff > 1$, and is thus in tension with BBN constraints on $Y_p$, unless the standard model neutrinos have a nonzero chemical potential.

The need to consider such exotic possibilities to possibly obviate the bounds we derive here demonstrates, once again, the important utility of cosmological observations on constraining fundamental neutrino particle physics. 

\begin{acknowledgments}
L. M. K. and T. D. J. acknowledge support from the DOE for this work;  C. L.  and T. D. J.  acknowledge the support of the NSF under Grant No. PHY-0854827. We thank the anonymous referee for helpful comments and suggestions.
\end{acknowledgments}

\appendix
\section*{Appendix: Analytical Approximation\label{appendix}}
  \renewcommand{\theequation}{A\arabic{equation}}
  \setcounter{equation}{0}  

Let us derive the approximate analytical result for $\Delta N^{\rm BBN}_{\rm eff}$,  Eqs. (\ref{analytic1}) and (\ref{analytic2}). 

\subsection{Oscillation amplitude in vacuum and in medium}

We consider  a basis  of five \n\ flavor states, ($\nu_\alpha$, $\alpha=e,\mu,\tau,s,r$), where $\nu_s, \nu_r$ are the sterile states.  These are related to the mass eigenstates by the mixing matrix $U$, $\nu_\alpha = \sum_{i=1,5} U_{\alpha i} \nu_i$,  where $U_{s 4} \simeq U_{r 5} \simeq 1$ and  $U$ is taken to be real for simplicity.  As stated in the main text, we use a number of assumptions to simplify the problem: (i) $U_{s5}=U_{r4}=0$, (ii)  $U_{\tau 4} = U_{\tau 5} = 0$, and (iii)  mass hierarchy $m_5 > m_4 \gg m_j$. 

Let us find the amplitude of active-sterile oscillations in vacuum, by calculating $P(\nua \rightarrow \nus)$ (for definiteness; analogous results hold for $\nur$), with $\alpha = e,\mu$. The channel $\nu_\tau  \rightarrow \nu_s$ has zero probability due to assumption (ii).  We use the standard notation $\Delta m^2_{ij}= m^2_i - m^2_j$. 

In all generality, the standard oscillation formalism gives 
\begin{equation}
P(\nua \rightarrow \nus) = -4 \sum_{i,j, i>j} U_{\alpha i} U_{s i} U_{\alpha j} U_{s j} \sin^2 \left( \frac{\Delta m^2_{ij}}{4 E} t \right)~. 
\end{equation}
Using assumption (i) and neglecting the lowest oscillation frequencies [assumption (iii)], we get
\beq
P(\nua \rightarrow \nus)&& \simeq  -4  U_{\alpha 4} U_{s 4} \sum_{i=1,2,3} U_{\alpha i} U_{s i} \sin^2 \left( \omega_s t \right) \nonumber \\
 && \simeq  4  U^2_{\alpha 4} U^2_{s 4}  \sin^2 \left( \omega_s t \right),
 \label{2nuappr}
\eeq
where $\omega_s = m^2_{4}/(4 E)$, and the last expression is obtained using the unitarity of the mixing matrix.   
Equation (\ref{2nuappr}) has the same form as the classic two-\n\ oscillation probability, with effective mixing 
\begin{equation}
\sin^2 2\theta_{\alpha s} \equiv 4  U^2_{\alpha 4} U^2_{s 4} \simeq 4  U^2_{\alpha 4}.
\label{effmix}
\end{equation}
An analogous result is obtained for $\nur$, with $\omega_r =m^2_{5}/(4 E)$ and $\sin^2 2\theta_{\alpha r}  \simeq 4  U^2_{\alpha 5}  $.
It is immediate to verify that, under the same assumptions as above, 
$P(\nua \rightarrow \nur) = P(\nur \rightarrow \nua)$  and $ P(\nua \rightarrow \nus) = P(\nus \rightarrow \nua)$.

Because of the thermal refraction potential,  the effective, two-neutrino oscillation amplitude is suppressed --- for both \ns\ and antineutrinos --- as follows (see, e.g., \cite{Foot:1996qc}):
\begin{eqnarray}
\sin^2 2\theta_m &\simeq& \frac{\sin^2 2\theta_{\alpha s}}{(1- b_\alpha(p,T))^2},\\
b_\alpha(p,T) &=& \frac{2 E \,V_\alpha}{\Delta m^2},
\label{mix}
\end{eqnarray}
where $V_\alpha$ is as in Eq.~(\ref{potential}) and we used $\cos 2\theta_{\alpha s} \simeq 1$, for convenience in the calculations that follow.   The expression (\ref{mix}) is valid for a CP-symmetric \n\ gas; the more complicated case with a lepton asymmetry will not be discussed here.

\subsection{Flavor evolution equation and its solution}

Let us now consider the production of the sterile \n\ $\nu_s$, using an effective two-neutrino system, $\nu_\alpha - \nu_s$ with the oscillation frequency and amplitude as outlined above. 
Let $f_s$ and $f_\alpha$ be the phase space distributions of $\nu_s$ and of one of the active species, and let $p$ be the neutrino momentum. 
We start with the evolution equation (see e.g., Foot and Volkas \cite{Foot:1996qc} and Dodelson and Widrow \cite{Dodelson:1993je})
\begin{eqnarray}
&&\left( \frac{\partial}{\partial t}  - H E \frac{\partial }{\partial E} \right) f_s(E,t)  = \nonumber\\
&& \sin^2 2\theta_m(E,t)  \frac{\Gamma_a(E,t)}{4}  \left(f_\alpha(E,t) - f_s(E,t) \right)~
\label{mastereq}
\end{eqnarray}
with  $\Gamma_\alpha$  being the collision rate,  Eq.~(\ref{collisionrate}).   Equation (\ref{mastereq}) is valid 
when the \n\ oscillation length is much shorter that the \n\ mean-free path, so that the effect of oscillations between two collisions is described by the averaged in-medium oscillation probability $\langle P \rangle = \sin^2 2\theta_m/2 $.  We have verified that this is always the case for our parameters of interest.\footnote{The hierarchy between the oscillation length and the mean free path is weak  for $T\gta 30$ MeV, and so one may doubt the accuracy of our equation.  In Ref.~\cite{Foot:1996qc} a more sophisticated equation is given, that does not rely on the hierarchy of lengths.  We have checked numerically  that this equation and Eq. (\ref{mastereq})  give very similar results. Therefore, we consider Eq. (\ref{mastereq}) for the sake of simplicity. }   We take $f_\alpha=f_\alpha(p/T)$ to be a Fermi-Dirac distribution, but the derivation in this section holds for any function of $p/T$.

Equation (\ref{mastereq}) can be simplified using \cite{Dodelson:1993je}
\beq
\left( \frac{\partial}{\partial t}  - H E \frac{\partial }{\partial E} \right)  = - H T \left.\frac{\partial }{\partial T} \right|_{y},  \qquad  y\equiv p/T ,
\eeq
and defining 
\begin{eqnarray}
x &\equiv&  \frac{2^{5/4}}{\pi M_W}\sqrt{A_\alpha \zeta(3) G_F} \frac{T^3}{m_4}\nonumber\\
&\simeq&3.53\times10^{-5} \sqrt{A_\alpha} \left(\frac{T}{\rm MeV}\right)^3 \left(\frac{\rm eV}{m_4}\right)
\end{eqnarray}
 so that $-b_\alpha(p,T)\equiv y^2 x^2 $.  
Thus, we have the new equation
\begin{equation} 
- H T \frac{\partial f_s}{\partial T}  = \frac{\sin^2 2\theta_{\alpha s} \Gamma_\alpha(y,x)}{2 (1 + x^2 y^2)^2} (f_\alpha(y) - f_s(x,y)),
\label{master2}
\end{equation}
where $y$ and $x$ should be treated as independent variables, and the differential equation should be solved with respect to $x$, with $y$ fixed. 
 
Changing the differentiation variable from $T$ to $x$ \cite{Dodelson:1993je}, and neglecting the temperature dependence of $g^\ast$, one finds a solution of the form
\beq
1- \frac{f_s}{f_\alpha} &&= \exp \left[ - \frac{m_4}{m_f}   U^2_{\alpha4} \frac{y_\alpha}{\sqrt{A_\alpha}}    \int^{\infty}_{xy} \frac{ d(x^\prime y)}{ (1 + (x^\prime y)^2)^2} \right] , \nonumber\\
m_f  
&& \simeq \frac{13 \sqrt{g^\ast} }{ G^{3/2}_F M_{\rm pl} M_W }  \simeq 1.05\times 10^{-3}~{\rm eV }~,
\eeq
where  $g^\ast =10$ has been used.  As discussed in Ref.~\cite{Dodelson:1993je}, if the lower integration limit is small, i.e. $xy\ll 1$ ($-b_\alpha(p,T)\ll 1$) at freeze-out, we can replace it with $0$,   for which the integral can be calculated easily, with $\pi/4$ as the result. In this limit, $f_s/f_\alpha$ is {\it independent} of $y$ \cite{Dodelson:1993je}, meaning that $f_s$ has the same spectral shape as $f_\alpha$, and the two only differ by an overall factor. 
For the freeze-out temperature $T_\nu\simeq 2$ MeV, and for $p = 3.15 T$, we get $xy\simeq 4\times 10^{-3} m_4/{\rm eV}$, so this condition holds for the range of masses of interest here. 

The final result for the ratio $f_s/f_\alpha$ is then
\begin{align}
\frac{f_s}{f_\alpha}= 1 - \exp \left[ - \frac{\pi}{4}\frac{m_4}{m_f}  U^2_{\alpha4} \frac{y_\alpha}{\sqrt{A_\alpha}}  \right].
\label{finalresult2nu}
\end{align}
In the limit $f_s/f_\alpha \ll 1$, an expansion of the exponential recovers the result in \cite{Dodelson:1993je,Melchiorri:2008gq}.

For our case, where two oscillation channels are present, $\nu_s \leftrightarrow \nu_e$ and $\nu_s \leftrightarrow \nu_\mu$, the generalization is immediate: 
\begin{align}
\frac{f_s}{f_\alpha} &=1 - \exp \left[ - \frac{\pi}{4}\frac{m_4}{m_f}  \left(U^2_{e 4} \frac{y_e}{\sqrt{A_e}} + U^2_{\mu 4} \frac{y_\mu}{\sqrt{A_\mu}} \right)  \right] \nonumber\\
&\simeq  1 - \exp \left[ \frac{-  2.06\times 10^3 }{\sqrt{g^\ast}}   \left(\frac{m_4 }{eV} \right)    \left( U^2_{e4} +  1.29 U^2_{\mu 4}  \right) \right] \nonumber\\
&\simeq  1 - \exp \left[ -  6.51\times 10^2    \left(\frac{m_4 }{eV} \right)    \left( U^2_{e4} +  1.29 U^2_{\mu 4}  \right) \right] .
\label{finalresultapp}
\end{align}
 A similar formula holds for the abundance of $\nu_r$, upon replacement of index: $4 \rightarrow 5$.
Note that the result in Eq. (\ref{finalresultapp}) can be rewritten in terms of a single, effective, $\nu_e - \nu_s$ system, with mixing angle $\sin^2 2\theta_{\rm eff} =4 ( U^2_{e4} +  1.29 U^2_{\mu 4}  ) $.  
\\

We find that our analytic solution, Eq.~(\ref{finalresultapp}), gives results around 10\% lower than  our numeric solution at points 1,2, and 3 in Table~\ref{parameters}. The main source of this discrepancy is $g^\ast$, which is kept fixed in the analytic solution, while its full temperature dependence is included in our numeric results. When $g^\ast$ is kept fixed in both calculations, results match to within 5\%.

\bibliography{mybiblio}

\begin{thebibliography}{57}%
\makeatletter
\providecommand \@ifxundefined [1]{%
 \@ifx{#1\undefined}
}%
\providecommand \@ifnum [1]{%
 \ifnum #1\expandafter \@firstoftwo
 \else \expandafter \@secondoftwo
 \fi
}%
\providecommand \@ifx [1]{%
 \ifx #1\expandafter \@firstoftwo
 \else \expandafter \@secondoftwo
 \fi
}%
\providecommand \natexlab [1]{#1}%
\providecommand \enquote  [1]{``#1''}%
\providecommand \bibnamefont  [1]{#1}%
\providecommand \bibfnamefont [1]{#1}%
\providecommand \citenamefont [1]{#1}%
\providecommand \href@noop [0]{\@secondoftwo}%
\providecommand \href [0]{\begingroup \@sanitize@url \@href}%
\providecommand \@href[1]{\@@startlink{#1}\@@href}%
\providecommand \@@href[1]{\endgroup#1\@@endlink}%
\providecommand \@sanitize@url [0]{\catcode `\\12\catcode `\$12\catcode
  `\&12\catcode `\#12\catcode `\^12\catcode `\_12\catcode `\%12\relax}%
\providecommand \@@startlink[1]{}%
\providecommand \@@endlink[0]{}%
\providecommand \url  [0]{\begingroup\@sanitize@url \@url }%
\providecommand \@url [1]{\endgroup\@href {#1}{\urlprefix }}%
\providecommand \urlprefix  [0]{URL }%
\providecommand \Eprint [0]{\href }%
\providecommand \doibase [0]{http://dx.doi.org/}%
\providecommand \selectlanguage [0]{\@gobble}%
\providecommand \bibinfo  [0]{\@secondoftwo}%
\providecommand \bibfield  [0]{\@secondoftwo}%
\providecommand \translation [1]{[#1]}%
\providecommand \BibitemOpen [0]{}%
\providecommand \bibitemStop [0]{}%
\providecommand \bibitemNoStop [0]{.\EOS\space}%
\providecommand \EOS [0]{\spacefactor3000\relax}%
\providecommand \BibitemShut  [1]{\csname bibitem#1\endcsname}%
\let\auto@bib@innerbib\@empty
\bibitem [{\citenamefont {Cowsik}\ and\ \citenamefont
  {McClelland}(1972)}]{Cowsik:1972gh}%
  \BibitemOpen
  \bibfield  {author} {\bibinfo {author} {\bibfnamefont {R.}~\bibnamefont
  {Cowsik}}\ and\ \bibinfo {author} {\bibfnamefont {J.}~\bibnamefont
  {McClelland}},\ }\href {\doibase 10.1103/PhysRevLett.29.669} {\bibfield
  {journal} {\bibinfo  {journal} {Phys.Rev.Lett.}\ }\textbf {\bibinfo {volume}
  {29}},\ \bibinfo {pages} {669} (\bibinfo {year} {1972})}\BibitemShut
  {NoStop}%
\bibitem [{\citenamefont {Steigman}\ \emph {et~al.}(1977)\citenamefont
  {Steigman}, \citenamefont {Schramm},\ and\ \citenamefont
  {Gunn}}]{Steigman:1977kc}%
  \BibitemOpen
  \bibfield  {author} {\bibinfo {author} {\bibfnamefont {G.}~\bibnamefont
  {Steigman}}, \bibinfo {author} {\bibfnamefont {D.~N.}\ \bibnamefont
  {Schramm}}, \ and\ \bibinfo {author} {\bibfnamefont {J.~E.}\ \bibnamefont
  {Gunn}},\ }\href {\doibase 10.1016/0370-2693(77)90176-9} {\bibfield
  {journal} {\bibinfo  {journal} {Phys. Lett.}\ }\textbf {\bibinfo {volume}
  {B66}},\ \bibinfo {pages} {202} (\bibinfo {year} {1977})}\BibitemShut
  {NoStop}%
\bibitem [{\citenamefont {{Krauss}}\ and\ \citenamefont
  {{Romanelli}}(1990)}]{1990ApJ...358...47K}%
  \BibitemOpen
  \bibfield  {author} {\bibinfo {author} {\bibfnamefont {L.~M.}\ \bibnamefont
  {{Krauss}}}\ and\ \bibinfo {author} {\bibfnamefont {P.}~\bibnamefont
  {{Romanelli}}},\ }\href {\doibase 10.1086/168962} {\bibfield  {journal}
  {\bibinfo  {journal} {\apj}\ }\textbf {\bibinfo {volume} {358}},\ \bibinfo
  {pages} {47} (\bibinfo {year} {1990})}\BibitemShut {NoStop}%
\bibitem [{\citenamefont {Kernan}\ and\ \citenamefont {Krauss}(1994)}]{k&k}%
  \BibitemOpen
  \bibfield  {author} {\bibinfo {author} {\bibfnamefont {P.~J.}\ \bibnamefont
  {Kernan}}\ and\ \bibinfo {author} {\bibfnamefont {L.~M.}\ \bibnamefont
  {Krauss}},\ }\href@noop {} {\bibfield  {journal} {\bibinfo  {journal} {Phys.
  Rev. Lett.}\ }\textbf {\bibinfo {volume} {72}},\ \bibinfo {pages} {3309}
  (\bibinfo {year} {1994})},\ \Eprint {http://arxiv.org/abs/astro-ph/9402010}
  {astro-ph/9402010} \BibitemShut {NoStop}%
\bibitem [{\citenamefont {Mangano}\ \emph {et~al.}(2005)\citenamefont {Mangano}
  \emph {et~al.}}]{mangano}%
  \BibitemOpen
  \bibfield  {author} {\bibinfo {author} {\bibfnamefont {G.}~\bibnamefont
  {Mangano}} \emph {et~al.},\ }\href {\doibase 10.1016/j.nuclphysb.2005.09.041}
  {\bibfield  {journal} {\bibinfo  {journal} {Nucl. Phys.}\ }\textbf {\bibinfo
  {volume} {B729}},\ \bibinfo {pages} {221} (\bibinfo {year} {2005})},\ \Eprint
  {http://arxiv.org/abs/hep-ph/0506164} {arXiv:hep-ph/0506164} \BibitemShut
  {NoStop}%
\bibitem [{\citenamefont {Krauss}\ \emph {et~al.}(2010)\citenamefont {Krauss},
  \citenamefont {Lunardini},\ and\ \citenamefont {Smith}}]{Krauss:2010xg}%
  \BibitemOpen
  \bibfield  {author} {\bibinfo {author} {\bibfnamefont {L.~M.}\ \bibnamefont
  {Krauss}}, \bibinfo {author} {\bibfnamefont {C.}~\bibnamefont {Lunardini}}, \
  and\ \bibinfo {author} {\bibfnamefont {C.}~\bibnamefont {Smith}},\
  }\href@noop {} {\  (\bibinfo {year} {2010})},\ \Eprint
  {http://arxiv.org/abs/1009.4666} {arXiv:1009.4666 [hep-ph]} \BibitemShut
  {NoStop}%
\bibitem [{\citenamefont {Hannestad}\ \emph {et~al.}(2012)\citenamefont
  {Hannestad}, \citenamefont {Tamborra},\ and\ \citenamefont
  {Tram}}]{Hannestad:2012ky}%
  \BibitemOpen
  \bibfield  {author} {\bibinfo {author} {\bibfnamefont {S.}~\bibnamefont
  {Hannestad}}, \bibinfo {author} {\bibfnamefont {I.}~\bibnamefont {Tamborra}},
  \ and\ \bibinfo {author} {\bibfnamefont {T.}~\bibnamefont {Tram}},\ }\href
  {\doibase 10.1088/1475-7516/2012/07/025} {\bibfield  {journal} {\bibinfo
  {journal} {JCAP}\ }\textbf {\bibinfo {volume} {1207}},\ \bibinfo {pages}
  {025} (\bibinfo {year} {2012})},\ \Eprint {http://arxiv.org/abs/1204.5861}
  {arXiv:1204.5861 [astro-ph.CO]} \BibitemShut {NoStop}%
\bibitem [{\citenamefont {Mirizzi}\ \emph {et~al.}(2012)\citenamefont
  {Mirizzi}, \citenamefont {Saviano}, \citenamefont {Miele},\ and\
  \citenamefont {Serpico}}]{Mirizzi:2012we}%
  \BibitemOpen
  \bibfield  {author} {\bibinfo {author} {\bibfnamefont {A.}~\bibnamefont
  {Mirizzi}}, \bibinfo {author} {\bibfnamefont {N.}~\bibnamefont {Saviano}},
  \bibinfo {author} {\bibfnamefont {G.}~\bibnamefont {Miele}}, \ and\ \bibinfo
  {author} {\bibfnamefont {P.~D.}\ \bibnamefont {Serpico}},\ }\href {\doibase
  10.1103/PhysRevD.86.053009} {\bibfield  {journal} {\bibinfo  {journal}
  {Phys.Rev.}\ }\textbf {\bibinfo {volume} {D86}},\ \bibinfo {pages} {053009}
  (\bibinfo {year} {2012})},\ \Eprint {http://arxiv.org/abs/1206.1046}
  {arXiv:1206.1046 [hep-ph]} \BibitemShut {NoStop}%
\bibitem [{\citenamefont {Ma}\ \emph {et~al.}(2000)\citenamefont {Ma},
  \citenamefont {Rajasekaran},\ and\ \citenamefont {Stancu}}]{Ma:1999im}%
  \BibitemOpen
  \bibfield  {author} {\bibinfo {author} {\bibfnamefont {E.}~\bibnamefont
  {Ma}}, \bibinfo {author} {\bibfnamefont {G.}~\bibnamefont {Rajasekaran}}, \
  and\ \bibinfo {author} {\bibfnamefont {I.}~\bibnamefont {Stancu}},\ }\href
  {\doibase 10.1103/PhysRevD.61.071302} {\bibfield  {journal} {\bibinfo
  {journal} {Phys.Rev.}\ }\textbf {\bibinfo {volume} {D61}},\ \bibinfo {pages}
  {071302} (\bibinfo {year} {2000})},\ \Eprint
  {http://arxiv.org/abs/hep-ph/9908489} {arXiv:hep-ph/9908489 [hep-ph]}
  \BibitemShut {NoStop}%
\bibitem [{\citenamefont {Palomares-Ruiz}\ \emph {et~al.}(2005)\citenamefont
  {Palomares-Ruiz}, \citenamefont {Pascoli},\ and\ \citenamefont
  {Schwetz}}]{PalomaresRuiz:2005vf}%
  \BibitemOpen
  \bibfield  {author} {\bibinfo {author} {\bibfnamefont {S.}~\bibnamefont
  {Palomares-Ruiz}}, \bibinfo {author} {\bibfnamefont {S.}~\bibnamefont
  {Pascoli}}, \ and\ \bibinfo {author} {\bibfnamefont {T.}~\bibnamefont
  {Schwetz}},\ }\href {\doibase 10.1088/1126-6708/2005/09/048} {\bibfield
  {journal} {\bibinfo  {journal} {JHEP}\ }\textbf {\bibinfo {volume} {0509}},\
  \bibinfo {pages} {048} (\bibinfo {year} {2005})},\ \Eprint
  {http://arxiv.org/abs/hep-ph/0505216} {arXiv:hep-ph/0505216 [hep-ph]}
  \BibitemShut {NoStop}%
\bibitem [{\citenamefont {Ichikawa}\ \emph {et~al.}(2007)\citenamefont
  {Ichikawa}, \citenamefont {Kawasaki}, \citenamefont {Nakayama}, \citenamefont
  {Senami},\ and\ \citenamefont {Takahashi}}]{Ichikawa:2007jv}%
  \BibitemOpen
  \bibfield  {author} {\bibinfo {author} {\bibfnamefont {K.}~\bibnamefont
  {Ichikawa}}, \bibinfo {author} {\bibfnamefont {M.}~\bibnamefont {Kawasaki}},
  \bibinfo {author} {\bibfnamefont {K.}~\bibnamefont {Nakayama}}, \bibinfo
  {author} {\bibfnamefont {M.}~\bibnamefont {Senami}}, \ and\ \bibinfo {author}
  {\bibfnamefont {F.}~\bibnamefont {Takahashi}},\ }\href {\doibase
  10.1088/1475-7516/2007/05/008} {\bibfield  {journal} {\bibinfo  {journal}
  {JCAP}\ }\textbf {\bibinfo {volume} {0705}},\ \bibinfo {pages} {008}
  (\bibinfo {year} {2007})},\ \Eprint {http://arxiv.org/abs/hep-ph/0703034}
  {arXiv:hep-ph/0703034 [HEP-PH]} \BibitemShut {NoStop}%
\bibitem [{\citenamefont {Cirelli}\ \emph {et~al.}(2005)\citenamefont
  {Cirelli}, \citenamefont {Marandella}, \citenamefont {Strumia},\ and\
  \citenamefont {Vissani}}]{Cirelli:2004cz}%
  \BibitemOpen
  \bibfield  {author} {\bibinfo {author} {\bibfnamefont {M.}~\bibnamefont
  {Cirelli}}, \bibinfo {author} {\bibfnamefont {G.}~\bibnamefont {Marandella}},
  \bibinfo {author} {\bibfnamefont {A.}~\bibnamefont {Strumia}}, \ and\
  \bibinfo {author} {\bibfnamefont {F.}~\bibnamefont {Vissani}},\ }\href
  {\doibase 10.1016/j.nuclphysb.2004.11.056} {\bibfield  {journal} {\bibinfo
  {journal} {Nucl. Phys.}\ }\textbf {\bibinfo {volume} {B708}},\ \bibinfo
  {pages} {215} (\bibinfo {year} {2005})},\ \Eprint
  {http://arxiv.org/abs/hep-ph/0403158} {arXiv:hep-ph/0403158} \BibitemShut
  {NoStop}%
\bibitem [{\citenamefont {Melchiorri}\ \emph {et~al.}(2009)\citenamefont
  {Melchiorri}, \citenamefont {Mena}, \citenamefont {Palomares-Ruiz},
  \citenamefont {Pascoli}, \citenamefont {Slosar} \emph
  {et~al.}}]{Melchiorri:2008gq}%
  \BibitemOpen
  \bibfield  {author} {\bibinfo {author} {\bibfnamefont {A.}~\bibnamefont
  {Melchiorri}}, \bibinfo {author} {\bibfnamefont {O.}~\bibnamefont {Mena}},
  \bibinfo {author} {\bibfnamefont {S.}~\bibnamefont {Palomares-Ruiz}},
  \bibinfo {author} {\bibfnamefont {S.}~\bibnamefont {Pascoli}}, \bibinfo
  {author} {\bibfnamefont {A.}~\bibnamefont {Slosar}},  \emph {et~al.},\ }\href
  {\doibase 10.1088/1475-7516/2009/01/036} {\bibfield  {journal} {\bibinfo
  {journal} {JCAP}\ }\textbf {\bibinfo {volume} {0901}},\ \bibinfo {pages}
  {036} (\bibinfo {year} {2009})},\ \Eprint {http://arxiv.org/abs/0810.5133}
  {arXiv:0810.5133 [hep-ph]} \BibitemShut {NoStop}%
\bibitem [{\citenamefont {Boyanovsky}\ and\ \citenamefont
  {Ho}(2007)}]{Boyanovsky:2006it}%
  \BibitemOpen
  \bibfield  {author} {\bibinfo {author} {\bibfnamefont {D.}~\bibnamefont
  {Boyanovsky}}\ and\ \bibinfo {author} {\bibfnamefont {C.~M.}\ \bibnamefont
  {Ho}},\ }\href@noop {} {\bibfield  {journal} {\bibinfo  {journal} {JHEP}\
  }\textbf {\bibinfo {volume} {07}},\ \bibinfo {pages} {030} (\bibinfo {year}
  {2007})},\ \Eprint {http://arxiv.org/abs/hep-ph/0612092}
  {arXiv:hep-ph/0612092} \BibitemShut {NoStop}%
\bibitem [{\citenamefont {Ho}\ and\ \citenamefont
  {Scherrer}(2013)}]{Ho:2012br}%
  \BibitemOpen
  \bibfield  {author} {\bibinfo {author} {\bibfnamefont {C.~M.}\ \bibnamefont
  {Ho}}\ and\ \bibinfo {author} {\bibfnamefont {R.~J.}\ \bibnamefont
  {Scherrer}},\ }\href@noop {} {\bibfield  {journal} {\bibinfo  {journal}
  {Phys. Rev. D 87,}\ }\textbf {\bibinfo {volume} {065016}} (\bibinfo {year}
  {2013})},\ \Eprint {http://arxiv.org/abs/1212.1689} {arXiv:1212.1689
  [hep-ph]} \BibitemShut {NoStop}%
\bibitem [{\citenamefont {Boehm}\ \emph {et~al.}(2012)\citenamefont {Boehm},
  \citenamefont {Dolan},\ and\ \citenamefont {McCabe}}]{Boehm:2012gr}%
  \BibitemOpen
  \bibfield  {author} {\bibinfo {author} {\bibfnamefont {C.}~\bibnamefont
  {Boehm}}, \bibinfo {author} {\bibfnamefont {M.~J.}\ \bibnamefont {Dolan}}, \
  and\ \bibinfo {author} {\bibfnamefont {C.}~\bibnamefont {McCabe}},\ }\href
  {\doibase 10.1088/1475-7516/2012/12/027} {\bibfield  {journal} {\bibinfo
  {journal} {JCAP}\ }\textbf {\bibinfo {volume} {1212}},\ \bibinfo {pages}
  {027} (\bibinfo {year} {2012})},\ \Eprint {http://arxiv.org/abs/1207.0497}
  {arXiv:1207.0497 [astro-ph.CO]} \BibitemShut {NoStop}%
\bibitem [{\citenamefont {Hooper}\ \emph {et~al.}(2012)\citenamefont {Hooper},
  \citenamefont {Queiroz},\ and\ \citenamefont {Gnedin}}]{Hooper:2011aj}%
  \BibitemOpen
  \bibfield  {author} {\bibinfo {author} {\bibfnamefont {D.}~\bibnamefont
  {Hooper}}, \bibinfo {author} {\bibfnamefont {F.~S.}\ \bibnamefont {Queiroz}},
  \ and\ \bibinfo {author} {\bibfnamefont {N.~Y.}\ \bibnamefont {Gnedin}},\
  }\href {\doibase 10.1103/PhysRevD.85.063513} {\bibfield  {journal} {\bibinfo
  {journal} {Phys.Rev.}\ }\textbf {\bibinfo {volume} {D85}},\ \bibinfo {pages}
  {063513} (\bibinfo {year} {2012})},\ \Eprint {http://arxiv.org/abs/1111.6599}
  {arXiv:1111.6599 [astro-ph.CO]} \BibitemShut {NoStop}%
\bibitem [{\citenamefont {Izotov}\ and\ \citenamefont {Thuan}(2010)}]{izotov}%
  \BibitemOpen
  \bibfield  {author} {\bibinfo {author} {\bibfnamefont {Y.~I.}\ \bibnamefont
  {Izotov}}\ and\ \bibinfo {author} {\bibfnamefont {T.~X.}\ \bibnamefont
  {Thuan}},\ }\href@noop {} {\bibfield  {journal} {\bibinfo  {journal}
  {Astrophys. J.}\ }\textbf {\bibinfo {volume} {710}},\ \bibinfo {pages} {L67}
  (\bibinfo {year} {2010})},\ \Eprint {http://arxiv.org/abs/1001.4440}
  {arXiv:1001.4440 [astro-ph.CO]} \BibitemShut {NoStop}%
\bibitem [{\citenamefont {Aver}\ \emph {et~al.}(2010)\citenamefont {Aver},
  \citenamefont {Olive},\ and\ \citenamefont {Skillman}}]{skillman}%
  \BibitemOpen
  \bibfield  {author} {\bibinfo {author} {\bibfnamefont {E.}~\bibnamefont
  {Aver}}, \bibinfo {author} {\bibfnamefont {K.~A.}\ \bibnamefont {Olive}}, \
  and\ \bibinfo {author} {\bibfnamefont {E.~D.}\ \bibnamefont {Skillman}},\
  }\href {\doibase 10.1088/1475-7516/2010/05/003} {\bibfield  {journal}
  {\bibinfo  {journal} {JCAP}\ }\textbf {\bibinfo {volume} {1005}},\ \bibinfo
  {pages} {003} (\bibinfo {year} {2010})},\ \Eprint
  {http://arxiv.org/abs/1001.5218} {arXiv:1001.5218 [astro-ph.CO]} \BibitemShut
  {NoStop}%
\bibitem [{\citenamefont {Aver}\ \emph {et~al.}(2012)\citenamefont {Aver},
  \citenamefont {Olive},\ and\ \citenamefont {Skillman}}]{Aver:2011bw}%
  \BibitemOpen
  \bibfield  {author} {\bibinfo {author} {\bibfnamefont {E.}~\bibnamefont
  {Aver}}, \bibinfo {author} {\bibfnamefont {K.~A.}\ \bibnamefont {Olive}}, \
  and\ \bibinfo {author} {\bibfnamefont {E.~D.}\ \bibnamefont {Skillman}},\
  }\href {\doibase 10.1088/1475-7516/2012/04/004} {\bibfield  {journal}
  {\bibinfo  {journal} {JCAP}\ }\textbf {\bibinfo {volume} {1204}},\ \bibinfo
  {pages} {004} (\bibinfo {year} {2012})},\ \Eprint
  {http://arxiv.org/abs/1112.3713} {arXiv:1112.3713 [astro-ph.CO]} \BibitemShut
  {NoStop}%
\bibitem [{\citenamefont {Mangano}\ and\ \citenamefont
  {Serpico}(2011)}]{Mangano:2011ar}%
  \BibitemOpen
  \bibfield  {author} {\bibinfo {author} {\bibfnamefont {G.}~\bibnamefont
  {Mangano}}\ and\ \bibinfo {author} {\bibfnamefont {P.~D.}\ \bibnamefont
  {Serpico}},\ }\href {\doibase 10.1016/j.physletb.2011.05.075} {\bibfield
  {journal} {\bibinfo  {journal} {Phys.Lett.}\ }\textbf {\bibinfo {volume}
  {B701}},\ \bibinfo {pages} {296} (\bibinfo {year} {2011})},\ \Eprint
  {http://arxiv.org/abs/1103.1261} {arXiv:1103.1261 [astro-ph.CO]} \BibitemShut
  {NoStop}%
\bibitem [{\citenamefont {Nollett}\ and\ \citenamefont
  {Holder}(2011)}]{Nollett:2011aa}%
  \BibitemOpen
  \bibfield  {author} {\bibinfo {author} {\bibfnamefont {K.~M.}\ \bibnamefont
  {Nollett}}\ and\ \bibinfo {author} {\bibfnamefont {G.~P.}\ \bibnamefont
  {Holder}},\ }\href@noop {} {\  (\bibinfo {year} {2011})},\ \Eprint
  {http://arxiv.org/abs/1112.2683} {arXiv:1112.2683 [astro-ph.CO]} \BibitemShut
  {NoStop}%
\bibitem [{\citenamefont {Hamann}\ \emph {et~al.}(2011)\citenamefont {Hamann},
  \citenamefont {Hannestad}, \citenamefont {Raffelt},\ and\ \citenamefont
  {Wong}}]{Hamann:2011ge}%
  \BibitemOpen
  \bibfield  {author} {\bibinfo {author} {\bibfnamefont {J.}~\bibnamefont
  {Hamann}}, \bibinfo {author} {\bibfnamefont {S.}~\bibnamefont {Hannestad}},
  \bibinfo {author} {\bibfnamefont {G.~G.}\ \bibnamefont {Raffelt}}, \ and\
  \bibinfo {author} {\bibfnamefont {Y.~Y.~Y.}\ \bibnamefont {Wong}},\ }\href
  {\doibase 10.1088/1475-7516/2011/09/034} {\bibfield  {journal} {\bibinfo
  {journal} {JCAP}\ }\textbf {\bibinfo {volume} {1109}},\ \bibinfo {pages}
  {034} (\bibinfo {year} {2011})},\ \Eprint {http://arxiv.org/abs/1108.4136}
  {arXiv:1108.4136 [astro-ph.CO]} \BibitemShut {NoStop}%
\bibitem [{\citenamefont {Ade}\ \emph {et~al.}(2013)\citenamefont {Ade} \emph
  {et~al.}}]{Ade:2013lta}%
  \BibitemOpen
  \bibfield  {author} {\bibinfo {author} {\bibfnamefont {P.}~\bibnamefont
  {Ade}} \emph {et~al.} (\bibinfo {collaboration} {Planck Collaboration}),\
  }\href@noop {} {\  (\bibinfo {year} {2013})},\ \Eprint
  {http://arxiv.org/abs/1303.5076} {arXiv:1303.5076 [astro-ph.CO]} \BibitemShut
  {NoStop}%
\bibitem [{\citenamefont {Hou}\ \emph {et~al.}(2012)\citenamefont {Hou},
  \citenamefont {Reichardt}, \citenamefont {Story}, \citenamefont {Follin},
  \citenamefont {Keisler} \emph {et~al.}}]{Hou:2012xq}%
  \BibitemOpen
  \bibfield  {author} {\bibinfo {author} {\bibfnamefont {Z.}~\bibnamefont
  {Hou}}, \bibinfo {author} {\bibfnamefont {C.}~\bibnamefont {Reichardt}},
  \bibinfo {author} {\bibfnamefont {K.}~\bibnamefont {Story}}, \bibinfo
  {author} {\bibfnamefont {B.}~\bibnamefont {Follin}}, \bibinfo {author}
  {\bibfnamefont {R.}~\bibnamefont {Keisler}},  \emph {et~al.},\ }\href@noop {}
  {\  (\bibinfo {year} {2012})},\ \Eprint {http://arxiv.org/abs/1212.6267}
  {arXiv:1212.6267 [astro-ph.CO]} \BibitemShut {NoStop}%
\bibitem [{\citenamefont {Hinshaw}\ \emph {et~al.}(2012)\citenamefont
  {Hinshaw}, \citenamefont {Larson}, \citenamefont {Komatsu}, \citenamefont
  {Spergel}, \citenamefont {Bennett} \emph {et~al.}}]{Hinshaw:2012fq}%
  \BibitemOpen
  \bibfield  {author} {\bibinfo {author} {\bibfnamefont {G.}~\bibnamefont
  {Hinshaw}}, \bibinfo {author} {\bibfnamefont {D.}~\bibnamefont {Larson}},
  \bibinfo {author} {\bibfnamefont {E.}~\bibnamefont {Komatsu}}, \bibinfo
  {author} {\bibfnamefont {D.}~\bibnamefont {Spergel}}, \bibinfo {author}
  {\bibfnamefont {C.}~\bibnamefont {Bennett}},  \emph {et~al.},\ }\href@noop {}
  {\  (\bibinfo {year} {2012})},\ \Eprint {http://arxiv.org/abs/1212.5226}
  {arXiv:1212.5226 [astro-ph.CO]} \BibitemShut {NoStop}%
\bibitem [{\citenamefont {Sievers}\ \emph {et~al.}(2013)\citenamefont
  {Sievers}, \citenamefont {Hlozek}, \citenamefont {Nolta}, \citenamefont
  {Acquaviva}, \citenamefont {Addison} \emph {et~al.}}]{Sievers:2013wk}%
  \BibitemOpen
  \bibfield  {author} {\bibinfo {author} {\bibfnamefont {J.~L.}\ \bibnamefont
  {Sievers}}, \bibinfo {author} {\bibfnamefont {R.~A.}\ \bibnamefont {Hlozek}},
  \bibinfo {author} {\bibfnamefont {M.~R.}\ \bibnamefont {Nolta}}, \bibinfo
  {author} {\bibfnamefont {V.}~\bibnamefont {Acquaviva}}, \bibinfo {author}
  {\bibfnamefont {G.~E.}\ \bibnamefont {Addison}},  \emph {et~al.},\
  }\href@noop {} {\  (\bibinfo {year} {2013})},\ \Eprint
  {http://arxiv.org/abs/1301.0824} {arXiv:1301.0824 [astro-ph.CO]} \BibitemShut
  {NoStop}%
\bibitem [{\citenamefont {Mueller}\ \emph {et~al.}(2011)\citenamefont
  {Mueller}, \citenamefont {Lhuillier}, \citenamefont {Fallot}, \citenamefont
  {Letourneau}, \citenamefont {Cormon} \emph {et~al.}}]{Mueller:2011nm}%
  \BibitemOpen
  \bibfield  {author} {\bibinfo {author} {\bibfnamefont {T.}~\bibnamefont
  {Mueller}}, \bibinfo {author} {\bibfnamefont {D.}~\bibnamefont {Lhuillier}},
  \bibinfo {author} {\bibfnamefont {M.}~\bibnamefont {Fallot}}, \bibinfo
  {author} {\bibfnamefont {A.}~\bibnamefont {Letourneau}}, \bibinfo {author}
  {\bibfnamefont {S.}~\bibnamefont {Cormon}},  \emph {et~al.},\ }\href
  {\doibase 10.1103/PhysRevC.83.054615} {\bibfield  {journal} {\bibinfo
  {journal} {Phys.Rev.}\ }\textbf {\bibinfo {volume} {C83}},\ \bibinfo {pages}
  {054615} (\bibinfo {year} {2011})},\ \Eprint {http://arxiv.org/abs/1101.2663}
  {arXiv:1101.2663 [hep-ex]} \BibitemShut {NoStop}%
\bibitem [{\citenamefont {Mention}\ \emph {et~al.}(2011)\citenamefont
  {Mention}, \citenamefont {Fechner}, \citenamefont {Lasserre}, \citenamefont
  {Mueller}, \citenamefont {Lhuillier} \emph {et~al.}}]{Mention:2011rk}%
  \BibitemOpen
  \bibfield  {author} {\bibinfo {author} {\bibfnamefont {G.}~\bibnamefont
  {Mention}}, \bibinfo {author} {\bibfnamefont {M.}~\bibnamefont {Fechner}},
  \bibinfo {author} {\bibfnamefont {T.}~\bibnamefont {Lasserre}}, \bibinfo
  {author} {\bibfnamefont {T.}~\bibnamefont {Mueller}}, \bibinfo {author}
  {\bibfnamefont {D.}~\bibnamefont {Lhuillier}},  \emph {et~al.},\ }\href
  {\doibase 10.1103/PhysRevD.83.073006} {\bibfield  {journal} {\bibinfo
  {journal} {Phys.Rev.}\ }\textbf {\bibinfo {volume} {D83}},\ \bibinfo {pages}
  {073006} (\bibinfo {year} {2011})},\ \Eprint {http://arxiv.org/abs/1101.2755}
  {arXiv:1101.2755 [hep-ex]} \BibitemShut {NoStop}%
\bibitem [{\citenamefont {Giunti}\ and\ \citenamefont
  {Laveder}(2011{\natexlab{a}})}]{Giunti:2010zu}%
  \BibitemOpen
  \bibfield  {author} {\bibinfo {author} {\bibfnamefont {C.}~\bibnamefont
  {Giunti}}\ and\ \bibinfo {author} {\bibfnamefont {M.}~\bibnamefont
  {Laveder}},\ }\href {\doibase 10.1103/PhysRevC.83.065504} {\bibfield
  {journal} {\bibinfo  {journal} {Phys.Rev.}\ }\textbf {\bibinfo {volume}
  {C83}},\ \bibinfo {pages} {065504} (\bibinfo {year} {2011}{\natexlab{a}})},\
  \Eprint {http://arxiv.org/abs/1006.3244} {arXiv:1006.3244 [hep-ph]}
  \BibitemShut {NoStop}%
\bibitem [{\citenamefont {Kaether}\ \emph {et~al.}(2010)\citenamefont
  {Kaether}, \citenamefont {Hampel}, \citenamefont {Heusser}, \citenamefont
  {Kiko},\ and\ \citenamefont {Kirsten}}]{Kaether:2010ag}%
  \BibitemOpen
  \bibfield  {author} {\bibinfo {author} {\bibfnamefont {F.}~\bibnamefont
  {Kaether}}, \bibinfo {author} {\bibfnamefont {W.}~\bibnamefont {Hampel}},
  \bibinfo {author} {\bibfnamefont {G.}~\bibnamefont {Heusser}}, \bibinfo
  {author} {\bibfnamefont {J.}~\bibnamefont {Kiko}}, \ and\ \bibinfo {author}
  {\bibfnamefont {T.}~\bibnamefont {Kirsten}},\ }\href {\doibase
  10.1016/j.physletb.2010.01.030} {\bibfield  {journal} {\bibinfo  {journal}
  {Phys.Lett.}\ }\textbf {\bibinfo {volume} {B685}},\ \bibinfo {pages} {47}
  (\bibinfo {year} {2010})},\ \Eprint {http://arxiv.org/abs/1001.2731}
  {arXiv:1001.2731 [hep-ex]} \BibitemShut {NoStop}%
\bibitem [{\citenamefont {Abdurashitov}\ \emph {et~al.}(2009)\citenamefont
  {Abdurashitov} \emph {et~al.}}]{Abdurashitov:2009tn}%
  \BibitemOpen
  \bibfield  {author} {\bibinfo {author} {\bibfnamefont {J.}~\bibnamefont
  {Abdurashitov}} \emph {et~al.} (\bibinfo {collaboration} {SAGE
  Collaboration}),\ }\href {\doibase 10.1103/PhysRevC.80.015807} {\bibfield
  {journal} {\bibinfo  {journal} {Phys.Rev.}\ }\textbf {\bibinfo {volume}
  {C80}},\ \bibinfo {pages} {015807} (\bibinfo {year} {2009})},\ \Eprint
  {http://arxiv.org/abs/0901.2200} {arXiv:0901.2200 [nucl-ex]} \BibitemShut
  {NoStop}%
\bibitem [{\citenamefont {Aguilar}\ \emph {et~al.}(2001)\citenamefont {Aguilar}
  \emph {et~al.}}]{Aguilar:2001ty}%
  \BibitemOpen
  \bibfield  {author} {\bibinfo {author} {\bibfnamefont {A.}~\bibnamefont
  {Aguilar}} \emph {et~al.} (\bibinfo {collaboration} {LSND}),\ }\href
  {\doibase 10.1103/PhysRevD.64.112007} {\bibfield  {journal} {\bibinfo
  {journal} {Phys. Rev.}\ }\textbf {\bibinfo {volume} {D64}},\ \bibinfo {pages}
  {112007} (\bibinfo {year} {2001})},\ \Eprint
  {http://arxiv.org/abs/hep-ex/0104049} {arXiv:hep-ex/0104049} \BibitemShut
  {NoStop}%
\bibitem [{\citenamefont {Aguilar-Arevalo}\ \emph
  {et~al.}(2009{\natexlab{a}})\citenamefont {Aguilar-Arevalo} \emph
  {et~al.}}]{AguilarArevalo:2008rc}%
  \BibitemOpen
  \bibfield  {author} {\bibinfo {author} {\bibfnamefont {A.}~\bibnamefont
  {Aguilar-Arevalo}} \emph {et~al.} (\bibinfo {collaboration} {MiniBooNE
  Collaboration}),\ }\href {\doibase 10.1103/PhysRevLett.102.101802} {\bibfield
   {journal} {\bibinfo  {journal} {Phys.Rev.Lett.}\ }\textbf {\bibinfo {volume}
  {102}},\ \bibinfo {pages} {101802} (\bibinfo {year} {2009}{\natexlab{a}})},\
  \Eprint {http://arxiv.org/abs/0812.2243} {arXiv:0812.2243 [hep-ex]}
  \BibitemShut {NoStop}%
\bibitem [{\citenamefont {Aguilar-Arevalo}\ \emph
  {et~al.}(2009{\natexlab{b}})\citenamefont {Aguilar-Arevalo} \emph
  {et~al.}}]{AguilarArevalo:2009xn}%
  \BibitemOpen
  \bibfield  {author} {\bibinfo {author} {\bibfnamefont {A.}~\bibnamefont
  {Aguilar-Arevalo}} \emph {et~al.} (\bibinfo {collaboration} {MiniBooNE
  Collaboration}),\ }\href {\doibase 10.1103/PhysRevLett.103.111801} {\bibfield
   {journal} {\bibinfo  {journal} {Phys.Rev.Lett.}\ }\textbf {\bibinfo {volume}
  {103}},\ \bibinfo {pages} {111801} (\bibinfo {year} {2009}{\natexlab{b}})},\
  \Eprint {http://arxiv.org/abs/0904.1958} {arXiv:0904.1958 [hep-ex]}
  \BibitemShut {NoStop}%
\bibitem [{\citenamefont {Aguilar-Arevalo}\ \emph {et~al.}(2010)\citenamefont
  {Aguilar-Arevalo} \emph {et~al.}}]{AguilarArevalo:2010wv}%
  \BibitemOpen
  \bibfield  {author} {\bibinfo {author} {\bibfnamefont {A.}~\bibnamefont
  {Aguilar-Arevalo}} \emph {et~al.} (\bibinfo {collaboration} {MiniBooNE
  Collaboration}),\ }\href {\doibase 10.1103/PhysRevLett.105.181801} {\bibfield
   {journal} {\bibinfo  {journal} {Phys.Rev.Lett.}\ }\textbf {\bibinfo {volume}
  {105}},\ \bibinfo {pages} {181801} (\bibinfo {year} {2010})},\ \Eprint
  {http://arxiv.org/abs/1007.1150} {arXiv:1007.1150 [hep-ex]} \BibitemShut
  {NoStop}%
\bibitem [{\citenamefont {Aguilar-Arevalo}\ \emph {et~al.}(2012)\citenamefont
  {Aguilar-Arevalo} \emph {et~al.}}]{AguilarArevalo:2012va}%
  \BibitemOpen
  \bibfield  {author} {\bibinfo {author} {\bibfnamefont {A.}~\bibnamefont
  {Aguilar-Arevalo}} \emph {et~al.} (\bibinfo {collaboration} {MiniBooNE
  Collaboration}),\ }\href@noop {} {\  (\bibinfo {year} {2012})},\ \Eprint
  {http://arxiv.org/abs/1207.4809} {arXiv:1207.4809 [hep-ex]} \BibitemShut
  {NoStop}%
\bibitem [{\citenamefont {Maltoni}\ and\ \citenamefont
  {Schwetz}(2007)}]{Maltoni:2007zf}%
  \BibitemOpen
  \bibfield  {author} {\bibinfo {author} {\bibfnamefont {M.}~\bibnamefont
  {Maltoni}}\ and\ \bibinfo {author} {\bibfnamefont {T.}~\bibnamefont
  {Schwetz}},\ }\href {\doibase 10.1103/PhysRevD.76.093005} {\bibfield
  {journal} {\bibinfo  {journal} {Phys.Rev.}\ }\textbf {\bibinfo {volume}
  {D76}},\ \bibinfo {pages} {093005} (\bibinfo {year} {2007})},\ \Eprint
  {http://arxiv.org/abs/0705.0107} {arXiv:0705.0107 [hep-ph]} \BibitemShut
  {NoStop}%
\bibitem [{\citenamefont {Kopp}\ \emph {et~al.}(2011)\citenamefont {Kopp},
  \citenamefont {Maltoni},\ and\ \citenamefont {Schwetz}}]{Kopp:2011qd}%
  \BibitemOpen
  \bibfield  {author} {\bibinfo {author} {\bibfnamefont {J.}~\bibnamefont
  {Kopp}}, \bibinfo {author} {\bibfnamefont {M.}~\bibnamefont {Maltoni}}, \
  and\ \bibinfo {author} {\bibfnamefont {T.}~\bibnamefont {Schwetz}},\ }\href
  {\doibase 10.1103/PhysRevLett.107.091801} {\bibfield  {journal} {\bibinfo
  {journal} {Phys.Rev.Lett.}\ }\textbf {\bibinfo {volume} {107}},\ \bibinfo
  {pages} {091801} (\bibinfo {year} {2011})},\ \Eprint
  {http://arxiv.org/abs/1103.4570} {arXiv:1103.4570 [hep-ph]} \BibitemShut
  {NoStop}%
\bibitem [{\citenamefont {Giunti}\ and\ \citenamefont
  {Laveder}(2011{\natexlab{b}})}]{Giunti:2011gz}%
  \BibitemOpen
  \bibfield  {author} {\bibinfo {author} {\bibfnamefont {C.}~\bibnamefont
  {Giunti}}\ and\ \bibinfo {author} {\bibfnamefont {M.}~\bibnamefont
  {Laveder}},\ }\href {\doibase 10.1103/PhysRevD.84.073008} {\bibfield
  {journal} {\bibinfo  {journal} {Phys.Rev.}\ }\textbf {\bibinfo {volume}
  {D84}},\ \bibinfo {pages} {073008} (\bibinfo {year} {2011}{\natexlab{b}})},\
  \Eprint {http://arxiv.org/abs/1107.1452} {arXiv:1107.1452 [hep-ph]}
  \BibitemShut {NoStop}%
\bibitem [{\citenamefont {Abazajian}\ \emph {et~al.}(2012)\citenamefont
  {Abazajian}, \citenamefont {Acero}, \citenamefont {Agarwalla}, \citenamefont
  {Aguilar-Arevalo}, \citenamefont {Albright} \emph
  {et~al.}}]{Abazajian:2012ys}%
  \BibitemOpen
  \bibfield  {author} {\bibinfo {author} {\bibfnamefont {K.}~\bibnamefont
  {Abazajian}}, \bibinfo {author} {\bibfnamefont {M.}~\bibnamefont {Acero}},
  \bibinfo {author} {\bibfnamefont {S.}~\bibnamefont {Agarwalla}}, \bibinfo
  {author} {\bibfnamefont {A.}~\bibnamefont {Aguilar-Arevalo}}, \bibinfo
  {author} {\bibfnamefont {C.}~\bibnamefont {Albright}},  \emph {et~al.},\
  }\href@noop {} {\  (\bibinfo {year} {2012})},\ \Eprint
  {http://arxiv.org/abs/1204.5379} {arXiv:1204.5379 [hep-ph]} \BibitemShut
  {NoStop}%
\bibitem [{\citenamefont {Conrad}\ \emph {et~al.}(2013)\citenamefont {Conrad},
  \citenamefont {Ignarra}, \citenamefont {Karagiorgi}, \citenamefont
  {Shaevitz},\ and\ \citenamefont {Spitz}}]{Conrad:2012qt}%
  \BibitemOpen
  \bibfield  {author} {\bibinfo {author} {\bibfnamefont {J.}~\bibnamefont
  {Conrad}}, \bibinfo {author} {\bibfnamefont {C.}~\bibnamefont {Ignarra}},
  \bibinfo {author} {\bibfnamefont {G.}~\bibnamefont {Karagiorgi}}, \bibinfo
  {author} {\bibfnamefont {M.}~\bibnamefont {Shaevitz}}, \ and\ \bibinfo
  {author} {\bibfnamefont {J.}~\bibnamefont {Spitz}},\ }\href {\doibase
  10.1155/2013/163897} {\bibfield  {journal} {\bibinfo  {journal} {Adv.High
  Energy Phys.}\ }\textbf {\bibinfo {volume} {2013}},\ \bibinfo {pages}
  {163897} (\bibinfo {year} {2013})},\ \Eprint {http://arxiv.org/abs/1207.4765}
  {arXiv:1207.4765 [hep-ex]} \BibitemShut {NoStop}%
\bibitem [{\citenamefont {Joudaki}\ \emph {et~al.}(2013)\citenamefont
  {Joudaki}, \citenamefont {Abazajian},\ and\ \citenamefont
  {Kaplinghat}}]{Joudaki:2012uk}%
  \BibitemOpen
  \bibfield  {author} {\bibinfo {author} {\bibfnamefont {S.}~\bibnamefont
  {Joudaki}}, \bibinfo {author} {\bibfnamefont {K.~N.}\ \bibnamefont
  {Abazajian}}, \ and\ \bibinfo {author} {\bibfnamefont {M.}~\bibnamefont
  {Kaplinghat}},\ }\href@noop {} {\bibfield  {journal} {\bibinfo  {journal}
  {Phys. Rev.}\ }\textbf {\bibinfo {volume} {D87}},\ \bibinfo {pages} {065003}
  (\bibinfo {year} {2013})},\ \Eprint {http://arxiv.org/abs/1208.4354}
  {arXiv:1208.4354 [astro-ph.CO]} \BibitemShut {NoStop}%
\bibitem [{\citenamefont {Archidiacono}\ \emph {et~al.}(2012)\citenamefont
  {Archidiacono}, \citenamefont {Fornengo}, \citenamefont {Giunti},\ and\
  \citenamefont {Melchiorri}}]{Archidiacono:2012ri}%
  \BibitemOpen
  \bibfield  {author} {\bibinfo {author} {\bibfnamefont {M.}~\bibnamefont
  {Archidiacono}}, \bibinfo {author} {\bibfnamefont {N.}~\bibnamefont
  {Fornengo}}, \bibinfo {author} {\bibfnamefont {C.}~\bibnamefont {Giunti}}, \
  and\ \bibinfo {author} {\bibfnamefont {A.}~\bibnamefont {Melchiorri}},\
  }\href {\doibase 10.1103/PhysRevD.86.065028} {\bibfield  {journal} {\bibinfo
  {journal} {Phys.Rev.}\ }\textbf {\bibinfo {volume} {D86}},\ \bibinfo {pages}
  {065028} (\bibinfo {year} {2012})},\ \Eprint {http://arxiv.org/abs/1207.6515}
  {arXiv:1207.6515 [astro-ph.CO]} \BibitemShut {NoStop}%
\bibitem [{\citenamefont {Enqvist}\ \emph {et~al.}(1992)\citenamefont
  {Enqvist}, \citenamefont {Kainulainen},\ and\ \citenamefont
  {Thomson}}]{Enqvist:1991qj}%
  \BibitemOpen
  \bibfield  {author} {\bibinfo {author} {\bibfnamefont {K.}~\bibnamefont
  {Enqvist}}, \bibinfo {author} {\bibfnamefont {K.}~\bibnamefont
  {Kainulainen}}, \ and\ \bibinfo {author} {\bibfnamefont {M.~J.}\ \bibnamefont
  {Thomson}},\ }\href {\doibase 10.1016/0550-3213(92)90442-E} {\bibfield
  {journal} {\bibinfo  {journal} {Nucl. Phys.}\ }\textbf {\bibinfo {volume}
  {B373}},\ \bibinfo {pages} {498} (\bibinfo {year} {1992})}\BibitemShut
  {NoStop}%
\bibitem [{\citenamefont {Riemer-Sorensen}\ \emph {et~al.}(2013)\citenamefont
  {Riemer-Sorensen}, \citenamefont {Parkinson}, \citenamefont {Davis},\ and\
  \citenamefont {Blake}}]{RiemerSorensen:2012ve}%
  \BibitemOpen
  \bibfield  {author} {\bibinfo {author} {\bibfnamefont {S.}~\bibnamefont
  {Riemer-Sorensen}}, \bibinfo {author} {\bibfnamefont {D.}~\bibnamefont
  {Parkinson}}, \bibinfo {author} {\bibfnamefont {T.~M.}\ \bibnamefont
  {Davis}}, \ and\ \bibinfo {author} {\bibfnamefont {C.}~\bibnamefont
  {Blake}},\ }\href {\doibase 10.1088/0004-637X/763/2/89} {\bibfield  {journal}
  {\bibinfo  {journal} {Astrophys.J.}\ }\textbf {\bibinfo {volume} {763}},\
  \bibinfo {pages} {89} (\bibinfo {year} {2013})},\ \Eprint
  {http://arxiv.org/abs/1210.2131} {arXiv:1210.2131 [astro-ph.CO]} \BibitemShut
  {NoStop}%
\bibitem [{\citenamefont {Giusarma}\ \emph {et~al.}(2013)\citenamefont
  {Giusarma}, \citenamefont {de~Putter},\ and\ \citenamefont
  {Mena}}]{Giusarma:2012ph}%
  \BibitemOpen
  \bibfield  {author} {\bibinfo {author} {\bibfnamefont {E.}~\bibnamefont
  {Giusarma}}, \bibinfo {author} {\bibfnamefont {R.}~\bibnamefont {de~Putter}},
  \ and\ \bibinfo {author} {\bibfnamefont {O.}~\bibnamefont {Mena}},\ }\href
  {\doibase 10.1103/PhysRevD.87.043515} {\bibfield  {journal} {\bibinfo
  {journal} {Phys.Rev.}\ }\textbf {\bibinfo {volume} {D87}},\ \bibinfo {pages}
  {043515} (\bibinfo {year} {2013})},\ \Eprint {http://arxiv.org/abs/1211.2154}
  {arXiv:1211.2154 [astro-ph.CO]} \BibitemShut {NoStop}%
\bibitem [{\citenamefont {Wang}\ \emph {et~al.}(2012)\citenamefont {Wang},
  \citenamefont {Meng}, \citenamefont {Zhang}, \citenamefont {Shan},
  \citenamefont {Gong} \emph {et~al.}}]{Wang:2012vh}%
  \BibitemOpen
  \bibfield  {author} {\bibinfo {author} {\bibfnamefont {X.}~\bibnamefont
  {Wang}}, \bibinfo {author} {\bibfnamefont {X.-L.}\ \bibnamefont {Meng}},
  \bibinfo {author} {\bibfnamefont {T.-J.}\ \bibnamefont {Zhang}}, \bibinfo
  {author} {\bibfnamefont {H.}~\bibnamefont {Shan}}, \bibinfo {author}
  {\bibfnamefont {Y.}~\bibnamefont {Gong}},  \emph {et~al.},\ }\href {\doibase
  10.1088/1475-7516/2012/11/018} {\bibfield  {journal} {\bibinfo  {journal}
  {JCAP}\ }\textbf {\bibinfo {volume} {1211}},\ \bibinfo {pages} {018}
  (\bibinfo {year} {2012})},\ \Eprint {http://arxiv.org/abs/1210.2136}
  {arXiv:1210.2136 [astro-ph.CO]} \BibitemShut {NoStop}%
\bibitem [{\citenamefont {Benson}\ \emph {et~al.}(2013)\citenamefont {Benson},
  \citenamefont {de~Haan}, \citenamefont {Dudley}, \citenamefont {Reichardt},
  \citenamefont {Aird} \emph {et~al.}}]{Benson:2011ut}%
  \BibitemOpen
  \bibfield  {author} {\bibinfo {author} {\bibfnamefont {B.}~\bibnamefont
  {Benson}}, \bibinfo {author} {\bibfnamefont {T.}~\bibnamefont {de~Haan}},
  \bibinfo {author} {\bibfnamefont {J.}~\bibnamefont {Dudley}}, \bibinfo
  {author} {\bibfnamefont {C.}~\bibnamefont {Reichardt}}, \bibinfo {author}
  {\bibfnamefont {K.}~\bibnamefont {Aird}},  \emph {et~al.},\ }\href@noop {}
  {\bibfield  {journal} {\bibinfo  {journal} {Apj, 763,}\ }\textbf {\bibinfo
  {volume} {147}} (\bibinfo {year} {2013})},\ \Eprint
  {http://arxiv.org/abs/1112.5435} {arXiv:1112.5435 [astro-ph.CO]} \BibitemShut
  {NoStop}%
\bibitem [{\citenamefont {Dodelson}\ \emph {et~al.}(2006)\citenamefont
  {Dodelson}, \citenamefont {Melchiorri},\ and\ \citenamefont
  {Slosar}}]{Dodelson:2005tp}%
  \BibitemOpen
  \bibfield  {author} {\bibinfo {author} {\bibfnamefont {S.}~\bibnamefont
  {Dodelson}}, \bibinfo {author} {\bibfnamefont {A.}~\bibnamefont
  {Melchiorri}}, \ and\ \bibinfo {author} {\bibfnamefont {A.}~\bibnamefont
  {Slosar}},\ }\href {\doibase 10.1103/PhysRevLett.97.041301} {\bibfield
  {journal} {\bibinfo  {journal} {Phys.Rev.Lett.}\ }\textbf {\bibinfo {volume}
  {97}},\ \bibinfo {pages} {041301} (\bibinfo {year} {2006})},\ \Eprint
  {http://arxiv.org/abs/astro-ph/0511500} {arXiv:astro-ph/0511500 [astro-ph]}
  \BibitemShut {NoStop}%
\bibitem [{\citenamefont {Birrell}\ \emph {et~al.}(2012)\citenamefont
  {Birrell}, \citenamefont {Yang}, \citenamefont {Chen},\ and\ \citenamefont
  {Rafelski}}]{Birrell:2012gg}%
  \BibitemOpen
  \bibfield  {author} {\bibinfo {author} {\bibfnamefont {J.}~\bibnamefont
  {Birrell}}, \bibinfo {author} {\bibfnamefont {C.-T.}\ \bibnamefont {Yang}},
  \bibinfo {author} {\bibfnamefont {P.}~\bibnamefont {Chen}}, \ and\ \bibinfo
  {author} {\bibfnamefont {J.}~\bibnamefont {Rafelski}},\ }\href@noop {} {\
  (\bibinfo {year} {2012})},\ \Eprint {http://arxiv.org/abs/1212.6943}
  {arXiv:1212.6943 [astro-ph.CO]} \BibitemShut {NoStop}%
\bibitem [{\citenamefont {Joudaki}\ and\ \citenamefont
  {Kaplinghat}(2012)}]{Joudaki:2011nw}%
  \BibitemOpen
  \bibfield  {author} {\bibinfo {author} {\bibfnamefont {S.}~\bibnamefont
  {Joudaki}}\ and\ \bibinfo {author} {\bibfnamefont {M.}~\bibnamefont
  {Kaplinghat}},\ }\href {\doibase 10.1103/PhysRevD.86.023526} {\bibfield
  {journal} {\bibinfo  {journal} {Phys.Rev.}\ }\textbf {\bibinfo {volume}
  {D86}},\ \bibinfo {pages} {023526} (\bibinfo {year} {2012})},\ \Eprint
  {http://arxiv.org/abs/1106.0299} {arXiv:1106.0299 [astro-ph.CO]} \BibitemShut
  {NoStop}%
\bibitem [{\citenamefont {Foot}\ and\ \citenamefont {Volkas}(1995)}]{FV}%
  \BibitemOpen
  \bibfield  {author} {\bibinfo {author} {\bibfnamefont {R.}~\bibnamefont
  {Foot}}\ and\ \bibinfo {author} {\bibfnamefont {R.~R.}\ \bibnamefont
  {Volkas}},\ }\href@noop {} {\bibfield  {journal} {\bibinfo  {journal} {Phys.\
  Rev.\ Lett.}\ }\textbf {\bibinfo {volume} {75}},\ \bibinfo {pages} {4350}
  (\bibinfo {year} {1995})}\BibitemShut {NoStop}%
\bibitem [{\citenamefont {Di~Bari}\ \emph {et~al.}(2000)\citenamefont
  {Di~Bari}, \citenamefont {Lipari},\ and\ \citenamefont
  {Lusignoli}}]{DiBari:1999ha}%
  \BibitemOpen
  \bibfield  {author} {\bibinfo {author} {\bibfnamefont {P.}~\bibnamefont
  {Di~Bari}}, \bibinfo {author} {\bibfnamefont {P.}~\bibnamefont {Lipari}}, \
  and\ \bibinfo {author} {\bibfnamefont {M.}~\bibnamefont {Lusignoli}},\ }\href
  {\doibase 10.1016/S0217-751X(00)00095-1} {\bibfield  {journal} {\bibinfo
  {journal} {Int.J.Mod.Phys.}\ }\textbf {\bibinfo {volume} {A15}},\ \bibinfo
  {pages} {2289} (\bibinfo {year} {2000})},\ \Eprint
  {http://arxiv.org/abs/hep-ph/9907548} {arXiv:hep-ph/9907548 [hep-ph]}
  \BibitemShut {NoStop}%
\bibitem [{\citenamefont {Abazajian}\ \emph {et~al.}(2001)\citenamefont
  {Abazajian}, \citenamefont {Fuller},\ and\ \citenamefont {Patel}}]{afp}%
  \BibitemOpen
  \bibfield  {author} {\bibinfo {author} {\bibfnamefont {K.}~\bibnamefont
  {Abazajian}}, \bibinfo {author} {\bibfnamefont {G.~M.}\ \bibnamefont
  {Fuller}}, \ and\ \bibinfo {author} {\bibfnamefont {M.}~\bibnamefont
  {Patel}},\ }\href {\doibase 10.1103/PhysRevD.64.023501} {\bibfield  {journal}
  {\bibinfo  {journal} {Phys. Rev.}\ }\textbf {\bibinfo {volume} {D64}},\
  \bibinfo {pages} {023501} (\bibinfo {year} {2001})},\ \Eprint
  {http://arxiv.org/abs/astro-ph/0101524} {arXiv:astro-ph/0101524} \BibitemShut
  {NoStop}%
\bibitem [{\citenamefont {Foot}\ and\ \citenamefont
  {Volkas}(1997)}]{Foot:1996qc}%
  \BibitemOpen
  \bibfield  {author} {\bibinfo {author} {\bibfnamefont {R.}~\bibnamefont
  {Foot}}\ and\ \bibinfo {author} {\bibfnamefont {R.~R.}\ \bibnamefont
  {Volkas}},\ }\href {\doibase 10.1103/PhysRevD.55.5147} {\bibfield  {journal}
  {\bibinfo  {journal} {Phys. Rev.}\ }\textbf {\bibinfo {volume} {D55}},\
  \bibinfo {pages} {5147} (\bibinfo {year} {1997})},\ \Eprint
  {http://arxiv.org/abs/hep-ph/9610229} {arXiv:hep-ph/9610229} \BibitemShut
  {NoStop}%
\bibitem [{\citenamefont {Dodelson}\ and\ \citenamefont
  {Widrow}(1994)}]{Dodelson:1993je}%
  \BibitemOpen
  \bibfield  {author} {\bibinfo {author} {\bibfnamefont {S.}~\bibnamefont
  {Dodelson}}\ and\ \bibinfo {author} {\bibfnamefont {L.~M.}\ \bibnamefont
  {Widrow}},\ }\href {\doibase 10.1103/PhysRevLett.72.17} {\bibfield  {journal}
  {\bibinfo  {journal} {Phys. Rev. Lett.}\ }\textbf {\bibinfo {volume} {72}},\
  \bibinfo {pages} {17} (\bibinfo {year} {1994})},\ \Eprint
  {http://arxiv.org/abs/hep-ph/9303287} {arXiv:hep-ph/9303287} \BibitemShut
  {NoStop}%
\end{thebibliography}%
\end{document}